\documentclass[preprint,review,,authoryear,12pt,3p]{elsarticle}
\usepackage{lineno,hyperref}

\journal{}

\usepackage{amsmath}									
\usepackage{float}
\usepackage{graphicx}
\usepackage{subfigure}
\usepackage{dsfont}
\usepackage{amssymb}
\usepackage{epstopdf}

\begin{document}

\begin{frontmatter}

\title{Motion of grain boundaries incorporating dislocation structure}

\author[mymainaddress]{Luchan Zhang}
\author[mymainaddress]{Yang Xiang\corref{mycorrespondingauthor}}
\cortext[mycorrespondingauthor]{Corresponding author}
\ead{maxiang@ust.hk}

\address[mymainaddress]{Department of Mathematics, The Hong Kong University of Science and Technology, Clear Water Bay, Kowloon, Hong Kong}

\begin{abstract}

In this paper, we present a continuum model for the dynamics of low angle grain boundaries in two dimensions based on the motion of constituent dislocations of the grain boundaries. The continuum model consists of an equation for the motion of grain boundaries (i.e., motion of the constituent dislocations in the grain boundary normal direction) and equations for the dislocation structure evolution on the grain boundaries.
This model is derived from the discrete dislocation dynamics model. The long-range elastic interaction between dislocations is included in the continuum model, which ensures that the dislocation structure on a grain boundary is consistent with the Frank's formula.
These evolutions of the grain boundary and its dislocation structure are able to describe both normal motion and tangential translation of the grain boundary and grain rotation due to both coupling and sliding.
Since the continuum model is based upon dislocation structure, it naturally accounts for the grain boundary shape change during the motion and rotation of the grain boundary by motion and reaction of the constituent dislocations.
  Using the derived continuum grain boundary dynamics model, simulations are performed for the dynamics of circular and non-circular two dimensional grain boundaries, and the results are validated by  discrete dislocation dynamics simulations.

\end{abstract}


\end{frontmatter}

\section{Introduction}
Grain boundaries are the interfaces of grains with different orientations and they play essential roles in the polycrystalline materials \citep{Sutton1995}. Grain boundaries migrate under various driving forces such as the capillarity force, the bulk energy difference, the concentration gradients across the boundary, and the applied stress field. The motion of grain boundaries crucially determines the mechanical and plastic behaviors of the materials.
The classical grain boundary dynamics models are based upon the motion by mean curvature to reduce the total interfacial energy \citep{Herring1951,Mullins1956,Sutton1995} using the misorientation-dependent grain boundary energy \citep{ReadShockley1950}. There are extensive studies in the literature on such motion of grain boundaries by using molecular dynamics or continuum simulations, e.g. \citep{Chenlq1994,Upmanyu1998,Kobayashi2000,Kazaryan2000,Upmanyu2002,Zhang2005,Upmanyu2006,Kirch2006,Selim2009,Srolovitz2010,Selim2016}.

It has been shown that the grain boundary normal motion can induce a coupled tangential motion which is proportional to the normal motion, as a result of the geometric constraint that the lattice planes must be continuous across the grain boundary \citep{Li1953223,Cahn20021,Cahn20044887}. Besides the tangential motion coupled with normal motion, there is another type of tangential motion that is the relative rigid-body translation of the grains along the boundary by sliding to reduce the grain boundary energy \citep{Li1962,Shewmon1966,Harris19982623,Kobayashi2000,Upmanyu2006,Selim2016}.  This sliding motion is independent of all other grain boundary motions. When a grain is embedded in another one, the tangential motions along a grain boundary give rise to a relative rotation between the two grains, leading to change of the misorientation of the grain boundary. In the grain rotation due to sliding, the misorientation angle goes to the nearby local energy minimum state (decreases for a low angle grain boundary), whereas in the grain rotation due to coupling, the misorientation angle increases. The coupling and sliding motions depend on the grain boundary structure and mechanisms of the dynamics. \citet{Rath2007} showed by a simple dislocation model and experimental observations that grain boundary motion does not have to couple with tangential motion. In fact, the coupling and sliding effects cancel out in their case.

Cahn and Taylor \citep{Cahn20044887,Taylor2007493} proposed a unified approach to the phenomena of the coupling and sliding associated with the grain boundary motion. They formulated the total tangential velocity $v_{\parallel}$ as the superposition of the coupling and sliding effects: $v_{\parallel}=\beta v_{\perp}+v_s$, where the tangential velocity induced by coupling effect is proportional the  normal velocity $v_{\perp}$ with the coupling parameter $\beta$, and $v_s$ is the tangential velocity produced by sliding effect. They discussed different cases for the rotation of a circular cylindrical grain embedded in another one \citep{Cahn20044887}.  When the grain does not have such symmetry, they proposed a generalized theory based on mass transfer by diffusion confined on the grain boundary \citep{Taylor2007493}.

 Molecular dynamics simulations have been performed to validate the theory of Cahn and Taylor on the coupling grain boundary motion to shear deformation for planar grain boundary \citep{Cahn20064953,Cahn20063965}, and grain boundary migration and grain rotation for closed circle cylindrical grain boundaries \citep{Cahn20021,Trautt20122407}.   Experimental observations have also been reported  on  the migration of low angle planar tilt boundaries coupled to shear deformation in Al bicrystal with stress \citep{Molodov20071843,Molodov20095396}. The ratios of the normal to the lateral motion that they measured are complied with the coupling factors in the theory and atomistic simulations by Cahn \textit{et al} \citep{Cahn20044887,Cahn20064953,Cahn20063965}. Phase field crystal model (an atomistic-level model) was employed  to simulate the dynamics of a two-dimensional circular grain, and grain rotation and translation by motion and reaction of the constituent dislocations were observed \citep{Wu2012407}. Phase field crystal simulations also showed that the coupling of grain boundary motion in polycrystalline systems can give rise to a rigid body translation of the lattice as a grain shrinks and that this process is mediated by dislocation climb and dislocation reactions \citep{Voorhees2016264}. Three-dimensional phase field crystal simulations were further performed to investigate the motion, rotation and dislocation reactions on a spherical grain in a BCC bicrystal \citep{Voorhees2017}.
Numerical simulations based upon the generalization of the Cahn-Taylor theory to noncircular grains  \citep{Taylor2007493} were performed using the level set method \citep{Gupta2014}.

 In this paper, we present a continuum model for the dynamics of low angle grain boundaries in two dimensions based on the motion of constituent dislocations of the grain boundaries. The continuum model consists of an equation for the motion of grain boundaries (i.e., motion of the constituent dislocations in the grain boundary normal direction) and equations for the dislocation structure evolution on the grain boundaries (Eqs.~\eqref{eqn:vn} and \eqref{eqn:vp}, or Eqs.~\eqref{eqn:vr} and \eqref{eqn:vpr} in Sec.~\ref{sec:model}).
 The long-range elastic interaction between dislocations is included in the continuum model, which ensures that the dislocation structures on the grain boundaries  are consistent with the Frank's formula for grain boundaries (the condition of cancellation of the far-field elastic fields).
These evolutions of the grain boundary and its dislocation structure are able to describe both normal motion and tangential translation of grain boundaries and grain rotation due to both coupling and sliding effects.
Since the continuum model is based upon dislocation structure, it naturally accounts for the grain boundary shape change during the motion and rotation of the grain boundary by motion and reaction of the constituent dislocations without explicit mass transfer. Unlike the  Cahn-Taylor theory \citep{Cahn20044887} in which the coupling effect is an assumption, our model is based on the motion of the grain boundary dislocations and the coupling effect is a result.  Our model also generalizes the  Cahn-Taylor theory
 by  incorporating detailed formulas of the driving forces for the normal and tangential grain boundary velocities that depend on the constituent dislocations, their Burgers vectors, and the grain boundary shape,
 as well as the  shape change of the grain boundaries. Our model is  different from their earlier generalization   based on the assumption of the coupling effect and mass transfer via surface diffusion \citep{Taylor2007493}.
 Note that in some existing continuum models for the motion of grain boundaries and grain rotation \citep{Li1962,Shewmon1966,Kobayashi2000,Upmanyu2006,Selim2016}, evolution of misorientation angle was included to reduce the grain boundary energy density.  These models able to capture the grain boundary sliding but not coupling. In our continuum model, we use dislocation densities on the grain boundary as variables instead of the misorientation angle, which enables the incorporation both the grain boundary coupling and sliding motions.
 Using the derived continuum grain boundary dynamics model, simulations are performed for the dynamics of circular and non-circular two dimensional grain boundaries.

 We also perform discrete dislocation dynamics simulations for the dynamics of these grain boundaries and the simulation results using the two models agree excellently with each other. In particular, both our continuum and discrete dislocation dynamics simulations show that without dislocation reaction, a non-circular grain boundary shrinks in a shape-preserving way due to the coupling effect, which is consistent with the prediction of the  continuum model in \citet{Taylor2007493} based on the assumption of the coupling effect and mass transfer via surface diffusion.

Our continuum grain boundary dynamics model is based upon the continuum framework for grain boundaries in \citet{Zhu2014175} derived rigorously from the discrete dislocation dynamics model.  Previously, a continuum model for the energy and dislocation structures on static grain boundaries has been developed \citep{Zhang2017} using this framework. In fact, the energetic and dynamic properties of grain boundaries  were understood
based on the underlying dislocation mechanisms in many of the available theories, simulations and experiments \citep{ReadShockley1950,Li1953223,Cahn20021,Cahn20044887,Cahn20064953,
Cahn20063965,Molodov20071843,Molodov20095396,Trautt20122407,Wu2012407,Voorhees2016264,Voorhees2017}.
Although some direct discrete dislocation dynamics simulations are able to provide detailed information of grain boundary or interface
structures and dynamics \citep{Lim2009,Quek2011,Lim2012}, continuum models of the dynamics of grain boundaries incorporating their dislocation structures are still desired for larger scale simulations without tracking individual dislocations.

The rest of the paper is organized as follows. In Sec.~\ref{sec:settings}, we describe the two dimensional settings of the grain boundaries with their dislocation structures, and review the continuum framework in \citet{Zhu2014175} based on which our continuum model will be developed.
In Sec.~\ref{sec:model}, we present our continuum model for the motion of grain boundaries. The variational derivation method of the continuum model is presented in \ref{sec:variation}. In Sec.~\ref{sec:Frank}, we discuss the calculation of the misorientation angle  based on the Frank's formula, which is maintained by the long-range dislocation interaction of the grain  boundary during its motion.
In Sec.~\ref{sec:compare}, we present the formula for the change of misorientation angle (grain rotation) derived from our continuum dynamics model. We also derive the formulas for the tangential motions of grain boundaries due to the coupling and sliding effects based on our continuum model, and make comparisons with the formulas in the Cahn-Taylor theory \citep{Cahn20044887,Trautt20122407}. Simulation results for the dynamics of circular and non-circular two dimensional grain boundaries using our continuum model  and comparisons with  discrete dislocation dynamics simulation results are presented in Sec.~\ref{sec:simulation}. Conclusions and discussion are made in Sec.~\ref{sec:conclusion}.

\section{Grain boundaries in two dimensions and review of the continuum framework in \citet{Zhu2014175}}\label{sec:settings}

We consider the two dimensional problem that one cylindrical grain is embedded in another grain with arbitrary cross-section shape. The inner grain has a misorientation angle $\theta$ relative to the outer grain, and the rotation axis is parallel to the cylindrical axis. The grain boundary is then a pure tilt boundary.

\begin{figure}[htbp]
\centering
    \includegraphics[width=.6\linewidth]{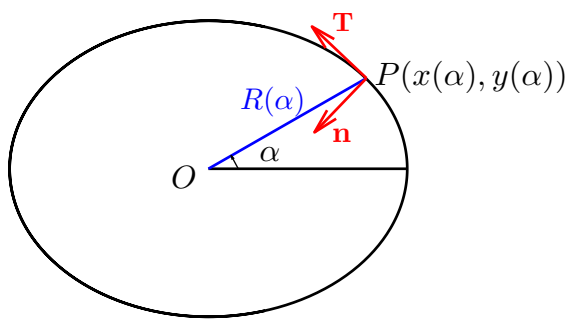}
    \caption{The cross-section of a cylindrical grain boundary with geometric center $O$. A point $P$ on the grain boundary can be written in polar coordinates as $(R(\alpha),\alpha)$,  where $R(\alpha)$ is the radius and $\alpha$ is the polar angle, and its Cartesian coordinate $\mathbf R(\alpha)= (x(\alpha),y(\alpha))=(R(\alpha)\cos\alpha,R(\alpha)\sin\alpha)$. }
    \label{fig:geometry}
\end{figure}

 We assume that
 the cross-section curve of the grain boundary $\Gamma$ is in the $xy$ plane and the rotation axis is in $z$ direction, and the geometric center (mass center) of the cross-section of the inner grain is the origin $O$ of the $xy$ plane.  We parameterize the two dimensional grain boundary $\Gamma$  by the polar angle $\alpha$, see  Fig.~\ref{fig:geometry}.
  All the functions defined on $\Gamma$,   such as the radius
  $R$ and the arclength parameter $s$ of the grain boundary,  are functions of the parameter $\alpha$. The derivative of a function $g$ defined on the grain boundary with respect to $\alpha$  is denoted by  $g'$, i.e.,  $g'=dg/d\alpha$.
   See \ref{sec:geometryformulas} for the formulas of the grain boundary tangent direction $\mathbf T$, the grain boundary normal direction $\mathbf n$,  the curvature of the grain boundary $\kappa$, and
  $ds/d\alpha$.

 Assume that on the grain boundary, there are $J$ dislocation arrays corresponding to $J$ different Burgers vectors $\mathbf{b}^{(j)}=(b_1^{(j)},b_2^{(j)},b_3^{(j)})$ with length $b^{(j)}=\|\mathbf b^{(j)}\|$ , $j=1,2,\cdots,J$, respectively.   All the dislocations are parallel to the $z$ axis, i.e., they are points in the $xy$ plane.

Our continuum model is based on the continuum framework proposed in \citet{Zhu2014175}. The dislocation densities on the grain boundaries are described by the dislocation density potential functions.
A dislocation density
potential function $\eta$ is a scalar function defined on the  grain boundary such that the constituent dislocations of the same Burgers vector are given by the integer-valued contour lines of $\eta$: {$\eta=i$, where $i$ is an integer}. The dislocation structure can be described in terms of the surface gradient of $\eta$ on the grain boundary, which is
$  \nabla_s \eta=\frac{d\eta}{ds}\mathbf T
 =\frac{{\eta}'(\alpha)}{\sqrt{{{R(\alpha)}^2+R'(\alpha)}^2}}\mathbf T$.
In particular, the local dislocation line direction is
$\mathbf t=\frac{\nabla_s \eta}{\|\nabla_s \eta\|} \times \mathbf n$,
and the inter-dislocation distance along the grain boundary is
$D=1/ \|\nabla_s \eta\|$.
Accordingly, the dislocation density per unit length on the grain boundary is
\begin{equation}\label{eqn:densityperlength}
\rho(\alpha)=\nabla_s \eta \cdot\mathbf T=\frac{d\eta}{ds}
 =\frac{{\eta}'(\alpha)}{\sqrt{{{R(\alpha)}^2+R'(\alpha)}^2}}.
\end{equation}
Here we allow the dislocation density $\rho$ to be negative to also include dislocations with opposite line directions.

In our continuum model, it is more convenient to use the dislocation density per unit polar angle, which is
\begin{equation}\label{eqn:density_angle}
\varrho(\alpha)=\rho(\alpha)\frac{ds}{d\alpha}=\rho(\alpha)\sqrt{R(\alpha)^2+R'(\alpha)^2}=\eta'(\alpha).
\end{equation}
Here we have used  Eqs.~\eqref{eqn:densityperlength} and \eqref{eqn:arclength}.
The surface gradient $\nabla_s \eta$ can be expressed in terms of $\varrho$ as
\begin{equation} \label{eqn:eta_by_rho}
\nabla_s\eta=\rho(\alpha) \mathbf T=\frac{\varrho(\alpha)}{\sqrt{R(\alpha)^2+R'(\alpha)^2}}\mathbf T.
\end{equation}
The dislocation line direction defined by $\mathbf t=\frac{\nabla_s \eta}{\|\nabla_s \eta\|} \times \mathbf n$ becomes
\begin{equation}\label{eqn:ddirection_rho}
\mathbf t=\frac{\varrho}{|\varrho|} \mathbf T\times \mathbf n.
 \end{equation}

 When there are $J$ dislocation arrays on the grain boundary, these constituent dislocations are represented by $\eta^{(j)},j=1,2,\cdots,J$, corresponding to $J$ different Burgers vectors $\mathbf{b}^{(j)}$, $j=1,2,\cdots,J$, respectively.

In the continuum framework presented in \citet{Zhu2014175}, the total energy associated with grain boundaries can be written as
\begin{equation}\label{eqn:totalenergy}
E=E_{\rm long}+E_{\rm local}+E_{\rm other},
\end{equation}
where $E_{\rm long}$ is the energy due to the long-range elastic interaction between the constituent dislocations of the grain boundaries,
$E_{\rm local}$  is  the line energy of the constituent dislocations corresponding the commonly used grain boundary energy in the literature \citep{Sutton1995,ReadShockley1950} and is a generalization of the classical Read-Shockley energy formula \citep{ReadShockley1950},
and  $E_{\rm other}$ includes the energy due to the interactions between the dislocation arrays and other stress fields such as the applied stress.

The variations of this total energy with respect to the change of the grain boundary and the change of dislocation structure on the grain boundary are
\begin{eqnarray}
\frac{\delta E}{\delta  r}&=&- \sum_{j=1}^J\|\nabla_s\eta^{(j)}\| \ \mathbf f_{\rm total}^{(j)}\cdot \mathbf n \label{eqn:var_n}, \ \
{\rm when} \ \delta\mathbf r = \mathbf n \delta  r, \vspace{1ex}
\\
\frac{\delta E}{\delta \eta^{(j)}}&=&\mathbf f_{\rm total}^{(j)}\cdot\frac{\nabla_s\eta^{(j)}}{\|\nabla_s\eta^{(j)}\|}, \ \
j=1,2,\cdots,J,\label{eqn:var_eta}
\end{eqnarray}
respectively, where
\begin{equation}
\mathbf f_{\rm total}^{(j)}={\mathbf f}_{\rm long}^{(j)}+{\mathbf f}_{\rm local}^{(j)}+{\mathbf f}_{\rm other}^{(j)}.\label{eqn:fs}
\end{equation}
Here   $\mathbf f_{\rm total}^{(j)}$ is the total force  on the $j$-th dislocation array, ${\mathbf f}_{\rm long}^{(j)}$ is the force that comes from the long-range interaction energy $E_{\rm long}$,
${\mathbf f}_{\rm local}^{(j)}$ is the force that comes from the local grain boundary energy $E_{\rm local}$,
 and ${\mathbf f}_{\rm other}^{(j)}$ is the force due to other stress fields (such as the applied stress field),  for $j=1,2,\cdots,J$.
 See Sec.~\ref{sec:model} for more discussion on these continuum forces.

Following the discrete dislocation dynamics model \citep{Xiang2003,Xiang2010,Zhu2014175},
 the continuum dynamics of these dislocation arrays along the grain boundary is given by
\begin{eqnarray}
\frac{\partial\eta^{(j)}}{\partial t}+\mathbf v^{(j)}\cdot \nabla_s\eta^{(j)}=0,\ \ \ \ j=1,2,\cdots,J,\label{eqn:evolution_eta}
\end{eqnarray}
where the dislocation velocity
$\mathbf v^{(j)}=\mathbf M_{\rm PK}^{(j)} \cdot \mathbf f^{(j)}_{\rm total}$,  and
  $\mathbf M_{\rm PK}^{(j)}$  is the mobility associated with the total Peach-Koehler force $\mathbf f^{(j)}_{\rm total}$  of the $j$-th dislocation array.
  If in the dynamics process, generation and removal of dislocations are critical, e.g.
 \citep{Li1953223,Shewmon1966,Cahn20021,Upmanyu2006,Wu2012407,Trautt20122407},  using Eq.~\eqref{eqn:var_eta},  the dislocation evolution equations are
\begin{eqnarray}
\frac{\partial \eta^{(j)}}{\partial t}=-m_j\frac{\delta E}{\delta \eta^{(j)}}
  =-m_j\left(\mathbf f^{(j)}\cdot\frac{\nabla_s\eta^{(j)}}{\|\nabla_s\eta^{(j)}\|}\right), \ \ \ \ j=1,2,\cdots,J,
\label{eqn:evolution3}
\end{eqnarray}
 where $\mathbf f^{(j)}$ is the total  force  on the $j$-th dislocation array in Eq.~\eqref{eqn:fs} or some of its contributions on the right-hand side, $m_j$ is some positive constant. The choice of Eqs.~\eqref{eqn:evolution_eta} and/or \eqref{eqn:evolution3} depends on the physics of the dynamics process. The motion of the grain boundary in general is, following Eq.~\eqref{eqn:var_n},
 \begin{eqnarray}
v_n=-m_n\frac{\delta E}{\delta r}
  =m_n \sum_{j=1}^J\|\nabla_s\eta^{(j)}\| \ \mathbf f_{\rm total}^{(j)}\cdot \mathbf n,
\label{eqn:evolution_S}
\end{eqnarray}
where $m_n$ is the mobility.

This continuum framework is general and applies to any dislocation arrays. Based on this framework, we will develop a two dimensional continuum model for the dynamics of grain boundaries including the motion and tangential translation of the grain boundaries and grain rotation that is consistent with  the discrete dislocation dynamics model and atomistic simulations, see the next section.
In the two dimensional problems, this new dislocation representation method based on dislocation density potential functions $\eta$'s and the classical method using the scalar dislocation densities are equivalent, see Eqs.~\eqref{eqn:density_angle} and \eqref{eqn:eta_by_rho}.  Continuum dynamics model for grain boundaries in three dimensions is being developed and will be presented elsewhere, in which it is more convenient to use  dislocation density potential functions for the structure of the constituent dislocations.

\section{The continuum model for grain boundary motion}\label{sec:model}

In this section, we present our continuum model for  grain boundary motion based on densities of the constituent dislocations. The grain boundary motion is coupled with the evolution of the constituent dislocations on the grain boundary, under both the long-range and local interactions of dislocations. The misorientation angle changes due to the coupling and sliding motions and the shape change of the grain boundary are naturally accommodated by the motion and reaction of the constituent dislocations.
  A critical idea to incorporate the coupling motion in the continuum model is to include the driving forces due to  energy variations under conservation of the constituent dislocations  \citep{Zhu2014175}, instead of the fixed grain boundary energy density as did in the literature. An example of such variational derivation is presented in \ref{sec:variation} for the two dimensional problem in terms of the contribution of the (local) grain boundary energy. The continuum model also incorporates the driving forces associated with energy variations with respect to the changes of dislocation densities  due to dislocation reactions or the applied stress on fixed grain boundary, leading to sliding motion of the grain boundary.

We consider the dynamics of a closed curved grain boundary $\Gamma$  with radius function $R(\alpha)$ and $J$ dislocation arrays on $\Gamma$  with Burgers vector $\mathbf{b}^{(j)}$, $j=1,2,\cdots,J$, as described in the previous section. The motion of the grain boundary and the evolution of the dislocation structure on it are described by
\begin{flalign}
v_n=&M_{\rm d}\sum_{j=1}^J\frac{|\varrho^{(j)}|}{\sum_{k=1}^J|\varrho^{(k)}|}\left({\mathbf f}_{\rm long}^{(j)}+{\mathbf f}_{\rm local}^{(j)}+{\mathbf f}_{\rm app}^{(j)}\right)\cdot \mathbf n +M_{\rm b}p,\label{eqn:vn}\\
\frac{\partial \varrho^{(j)}}{\partial t}=&-\left( \frac{\varrho^{(j)}}{\sqrt{R^2+R'^2}} M_{\rm d} {\mathbf f}_{\rm long}^{(j)}\cdot \mathbf T\right)'
-M_{\rm t}\frac{\partial \gamma}{\partial \varrho^{(j)}}
-M_{\rm a}\left( \frac{\varrho^{(j)}}{|\varrho^{(j)}|} \ {\mathbf f}_{\rm app}^{(j)}\cdot\mathbf T\right)', \label{eqn:vp}\\ &j=1,2,\cdots,J. \nonumber
\end{flalign}
Note that with the grain boundary normal velocity in Eq.~\eqref{eqn:vn},  Eq.~\eqref{eqn:vp} holds when every point on the grain boundary moves in the normal direction of the grain boundary.

In the simulations of a finite grain embedded in another one, instead of using the grain boundary normal velocity $v_n$ in Eq.~\eqref{eqn:vn} directly, it is more convenient to move each point on the grain boundary in the radial direction:
\begin{flalign}
\frac{\partial R}{\partial t}=&-v_n/\cos\lambda,\label{eqn:vr}\\
\frac{\partial \varrho^{(j)}}{\partial t}=&-\left[ \frac{\varrho^{(j)}}{\sqrt{R^2+R'^2}} \left(M_{\rm d} {\mathbf f}_{\rm long}^{(j)}\cdot \mathbf T
+v_n\tan\lambda\right)\right]'-M_{\rm t}\frac{\partial \gamma}{\partial \varrho^{(j)}}
-M_{\rm a}\left( \frac{\varrho^{(j)}}{|\varrho^{(j)}|} \ {\mathbf f}_{\rm app}^{(j)}\cdot\mathbf T\right)',\label{eqn:vpr}\\
&\ \ \ j=1,2,\cdots,J, \nonumber
\end{flalign}
where $\lambda$ is the angle between the normal direction of the grain boundary and its inward radial direction (see Fig.~\ref{fig:geometry}), and
$\cos\lambda=R/\sqrt{R^2+R'^2}$, $\sin\lambda=R'/\sqrt{R^2+R'^2}$.
With Eq.~\eqref{eqn:vr}, Eq.~\eqref{eqn:vpr} holds when every point on the grain boundary moves in the radial direction of the grain boundary, i.e., the value of the parameter $\alpha$ for each material point on the grain boundary does not change.

Eq.~\eqref{eqn:vn} gives  the normal velocity of the grain boundary $v_n$, which represents the motion of the constituent dislocations in the normal direction of the grain boundary.
In Eq.~\eqref{eqn:vn}, ${\mathbf f}_{\rm long}^{(j)}$, ${\mathbf f}_{\rm local}^{(j)}$ and ${\mathbf f}_{\rm app}^{(j)}$ are the continuum long-range dislocation interaction force, the continuum local force, and the force due to applied stress field, respectively, on a dislocation with Burgers vector  $\mathbf{b}^{(j)}$, and $M_{\rm d}$ is the mobility of the dislocations. The velocity of a point on the grain boundary is the weighted average of the velocities of dislocations on the grain boundary with different Burgers vectors. This modification has been made from the variational normal velocity formula in Eq.~\eqref{eqn:evolution_S} so that the velocity of the grain boundary is  consistent with the dynamics of its constituent dislocations \citep{Cahn20044887}.
Recall that the available  atomistic simulations \citep{Cahn20021} and phase field crystal simulations \citep{Wu2012407,Voorhees2016264,Voorhees2017} were performed at high temperatures to examine the coupling motion of curved grain boundaries, which is purely geometric \citep{Cahn20044887}. At these temperatures, the dislocation climb mobility is comparable with that of dislocation glide.
Moreover, the atomistic simulations of \citet{Trautt20122407} showed that the grain boundary dislocations do not necessarily move by dislocation climb, which is in general required (simply based on the dislocation motion direction)  for the motion of curved low angle grain boundaries and is very small at  a not very high temperature; instead, the dislocations move by a chain of dissociation/association processes. Such dislocation reaction mechanism leads to an effectively large mobility of the grain boundary dislocations in their climb direction. Following these simulation results,
 we assume that the dislocation  mobility $M_{\rm d}$ is isotropic in our continuum model to fully examine the purely geometric coupling motion of low angle grain boundaries.
Influences of the dislocation glide, climb and cross-slip mobilities and dislocation dissociation/association reactions on the grain boundary motion can be further incorporated in the continuum model by using different mobilities for these dislocation motions. These will be explored in the future work.
Another driving force for the grain boundary motion is the difference between the bulk energy densities of the two grains denoted by $p$ \citep{Cahn20044887,Trautt20122407}, and $M_{\rm b}$ is the grain boundary mobility associated with this driving force.

Eq.~\eqref{eqn:vp} describes the evolution of the dislocation structure on the grain boundary. (Recall that $g'=dg/d\alpha$ for a function $g(\alpha)$.) These equations are based on the dislocation motion along the grain boundary, driven by the continuum long-range interaction force, dislocation reaction, and the applied stress field.
The first term in Eq.~\eqref{eqn:vp} is the motion of dislocations along the grain boundary following conservation law in Eq.~\eqref{eqn:evolution_eta} driven by the continuum long-range elastic force. The importance of this term is to maintain the dislocation structure and the condition of cancellation of the far-field stress fields for the grain boundary (or the Frank's formula \citep{Frank,Bilby}) in a way that is consistent with the discrete dislocation dynamics.

The second term in Eq.~\eqref{eqn:vp} is the driving force due to variation of the grain boundary energy density  $\gamma$ (when the long-range elastic interaction vanishes)  with respect to the change of dislocation density $\varrho$ on the fixed grain boundary.
The third term in Eq.~\eqref{eqn:vp} is due to the effect of the applied stress.
Since  the equilibrium dislocation structure on a grain boundary is stable except for the change of the misorientation angle (i.e. sliding along a fixed grain boundary) \citep{Xiang-Yan2017},
the major influences of the local energy and the applied stress on a fixed grain boundary are to change the dislocation structure by dislocation reactions, leading to grain boundary sliding \citep{Li1953223,Shewmon1966,Cahn20021,Upmanyu2006,Wu2012407,Trautt20122407}, see Sec.~\ref{sec:rotation} for more discussion.
 In these processes, the dislocations on the grain boundary are not conserved and the last two terms in Eq.~\eqref{eqn:vp} account for such changes of dislocation structure due to these two driving forces (see also Eq.~\eqref{eqn:evolution3}), in which  $M_{\rm t}$ and $M_{\rm a}$ are the  mobilities associated with the driving forces of the local energy and the applied stress, respectively. Note that these grain boundary dislocation reactions are not energetically favorable and energy barriers have to be overcome during the reaction processes \citep{Cahn20021,Trautt20122407}.  The rates of these dislocation reactions are reflected in the mobilities $M_{\rm t}$ and $M_{\rm a}$.

 In these evolution equations,  ${\mathbf f}_{\rm long}^{(j)}$ is the force that comes from the long-range interaction energy $E_{\rm long}$ in Eq.~\eqref{eqn:totalenergy}.
 This continuum long-range interaction force on a dislocation  of the $j$-th dislocation arrays located at the point $(x,y)$ is
 \begin{eqnarray}\label{eqn:flong0}
{\mathbf f}_{\rm long}^{(j)}(x,y)=\sum_{k=1}^J  \int_\Gamma\mathbf{\tilde{f}}^{(j,k)}(x,y;x_1,y_1)ds,
\end{eqnarray}
where $ds$ is the line element of the integral along the grain boundary $\Gamma$, the point $(x_1,y_1)$ varies along $\Gamma$ in the integral, and $\int_\Gamma\mathbf{\tilde{f}}^{(j,k)}(x,y;x_1,y_1)ds$ is the force acting on a dislocation  of the $j$-th dislocation array at the point $(x,y)$ generated by the $k$-th dislocation array, with
\begin{flalign}
\mathbf{\tilde{f}}^{(j,k)}&=(\tilde{f}_1^{(j,k)}, \tilde{f}_2^{(j,k)}),
\end{flalign}
\begin{flalign}
\tilde{f}_1^{(j,k)}(x,y;x_1,y_1) &={\textstyle\frac{\mu}{2\pi(1-\nu)}\frac{1}{[(x-x_1)^2+(y-y_1)^2]^2}\frac{\varrho^{(k)}(x_1,y_1)}{\sqrt{{x'_1}^2+{y'_1}^2}}
\frac{\varrho^{(j)}(x,y)}{|\varrho^{(j)}(x,y)|}}\vspace{1ex}\nonumber\\
& \cdot \left\{  [(x-x_1)^3-(x-x_1)(y-y_1)^2]b_1^{(k)}b_1^{(j)}\right.\vspace{1ex}\nonumber\\
&+[(x-x_1)^2(y-y_1) -(y-y_1)^3]b_2^{(k)}b_1^{(j)} \vspace{1ex}\nonumber\\
&+   [(x-x_1)^2(y-y_1)- (y-y_1)^3]b_1^{(k)}b_2^{(j)} \vspace{1ex}\nonumber\\
&\left.+ [(x-x_1)^3+3(x-x_1)(y-y_1)^2]b_2^{(k)} b_2^{(j)} \right\},
\end{flalign}
\begin{flalign}
\tilde{f}_2^{(j,k)}(x,y;x_1,y_1) &={\textstyle \frac{\mu}{2\pi(1-\nu)}\frac{1}{((x-x_1)^2+(y-y_1)^2)^2}
\frac{\varrho^{(k)}(x_1,y_1)}{\sqrt{{x'_1}^2+{y'_1}^2}}
\frac{\varrho^{(j)}(x,y)}{|\varrho^{(j)}(x,y)|}}\vspace{1ex} \nonumber \\
&\cdot \left\{  [3(x-x_1)^2(y-y_1)+(y-y_1)^3]b_1^{(k)} b_1^{(j)}\right.\vspace{1ex}\nonumber\\
&+ [-(x-x_1)^3+ (x-x_1)(y-y_1)^2]b_2^{(k)}b_1^{(j)}\vspace{1ex} \nonumber\\
& + [-(x-x_1)^3 +(x-x_1)(y-y_1)^2]b_1^{(k)}b_2^{(j)} \vspace{1ex} \nonumber\\
&\left.+ [-(x-x_1)^2(y-y_1) + (y-y_1)^3]b_2^{(k)}b_2^{(j)}\right\}.\label{eqn:flong1}
\end{flalign}
Here $\mu$ is the shear modulus and $\nu$ is the Poisson ratio.
 The integral in Eq.~\eqref{eqn:flong0} is over all the grain boundaries if there are multiple grain boundaries in the system. The long-range interaction force ${\mathbf f}_{\rm long}$ and the long-range interaction energy $E_{\rm long}$ vanish for an equilibrium grain boundary \citep{Sutton1995,HL,Zhu2014175}.

The continuum local force on a dislocation of the $j$-th dislocation arrays in these evolution equations is
\begin{eqnarray}\label{eqn:local_force}
{\mathbf f}_{\text{local}}^{(j)}\cdot\mathbf{n}=\frac{\mu (b^{(j)})^2}{4\pi(1-\nu)}\kappa,
\end{eqnarray}
where $\kappa$ is the curvature of the  grain boundary. This local force comes from the grain boundary $E_{\rm local}$ (the line energy of
the constituent dislocations)   in Eq.~\eqref{eqn:totalenergy}. The energy density of $E_{\rm local}$ is \citep{Zhu2014175,Zhang2017}
\begin{flalign}
\gamma={\displaystyle \sum_{j=1}^J  \frac{\mu(b^{(j)})^2}{4\pi(1-\nu)}\!\left[1-\nu\frac{(\mathbf T\! \times \!\mathbf{n} \!\cdot\! \mathbf{b}^{(j)})^2}{(b^{(j)})^2 }\right]\! \frac{|\varrho^{(j)}|}{\sqrt{R^2+R'^2}}
\log\! \frac{1}{r_g\sqrt{{\varrho^{(j)}}^2/(R^2+R'^2)  +\epsilon}}},\label{eqn:gb_density}
\end{flalign}
where $r_g$ is a dislocation core parameter and $\epsilon$ is a numerical cutoff parameter.
This is a generalization of the classical Read-Shockley energy formula \citep{ReadShockley1950}.

These force formulas are the two-dimensional forms of the continuum forces on  low angle grain boundaries  derived by \citet{Zhu2014175}  based on the discrete dislocation model \citep{HL}.
It has been shown that the summation of the continuum long-range and local forces provides a good approximation to the Peach-Koehler force on a dislocation in the discrete dislocation dynamics model.

The Peach-Koehler force due to the applied stress $\pmb\sigma_{\rm app}$ is
${\mathbf f}_{\rm app}^{(j)}=(\pmb\sigma_{\rm app}\cdot \mathbf{b}^{(j)})\times{\mathbf t}^{(j)}$,
where ${\mathbf t}^{(j)}$ is the line direction of a dislocation in the $j$-th dislocation array on the grain boundary.
Using Eq.~\eqref{eqn:ddirection_rho}, the contribution of the applied stress the grain boundary velocity given in Eq.~\eqref{eqn:vn} can be written as
\begin{eqnarray}\label{eqn:fapp_dot_n}
{\mathbf f}_{\rm app}^{(j)}\cdot \mathbf n
= -\frac{\varrho^{(j)}}{|\varrho^{(j)}|} (\mathbf{b}^{(j)}\cdot \pmb\sigma_{\rm app}\cdot \mathbf T),
\end{eqnarray}
and the contribution of the applied stress to the dislocation structure evolution equation \eqref{eqn:vp} can be written as
\begin{flalign}\label{eqn:contributionofapplied}
\left(\frac{\varrho^{(j)}}{|\varrho^{(j)}|} \ {\mathbf f}_{\rm app}^{(j)}\cdot\mathbf T\right)'
=\left(\mathbf{b}^{(j)}\cdot \pmb\sigma_{\rm app}\cdot \mathbf n\right)'=-\kappa\sqrt{R^2+R'^2} \ (\mathbf{b}^{(j)}\cdot \pmb\sigma_{\rm app}\cdot \mathbf T),
\end{flalign}
where the last expression holds for a constant applied stress $\pmb\sigma_{\rm app}$, otherwise the expression in the middle should be used.

{\bf Remark:} Essentially, it is equivalent to use the dislocation density per unit length $\rho^{(j)}$, $j=1,2,\cdots,J$, as the variables in the continuum grain boundary dynamics model in Eqs.~\eqref{eqn:vn} and \eqref{eqn:vp}. However, such evolution equations of  $\rho^{(j)}$'s will no longer be as simple as Eq.~\eqref{eqn:vp} because the arclength of the grain boundary also evolves as the grain boundary migrates.

\section{Misorientation angle and effect of long-range dislocation interaction}\label{sec:Frank}

In our continuum model, the misorientation angle $\theta$ between  the two grains can be calculated based on the Frank's formula, which is maintained during the motion of the grain boundary by the long-range elastic interaction between the constituent dislocations of the grain  boundary.

\subsection{Frank's formula and misorientation angle}
With the continuum model in Eqs.~\eqref{eqn:vn} and \eqref{eqn:vp} (or Eqs.~\eqref{eqn:vr} and \eqref{eqn:vpr}) for the motion of the grain boundary and evolution of the dislocation structure on it, the misorientation angle $\theta$ between the two grains at any point on the grain boundary can be calculated  based on the Frank's formula  \citep{Frank,Bilby}, which is a condition for an equilibrium grain boundary dislocation structure  and is equivalent to the cancellation of the long-range elastic fields  \citep{Frank,Bilby,Zhu2014175}. Using the representation of dislocation density potential functions, the Frank's formula is  \citep{Zhu2014175}
\begin{flalign}
\theta(\mathbf v\times \mathbf a)- \sum_{j=1}^J \mathbf b^{(j)}(\nabla_s\eta^{(j)}\cdot\mathbf v)=0,
\end{flalign}
where $\mathbf a$ is the rotation axis and $\mathbf v$ is any vector tangent to the grain boundary. In the case of two dimensional tilt boundaries being considered in this paper, the rotation axis $\mathbf a$ is in the $+z$ direction, and the Frank's formula can be written as
\begin{flalign}\label{eqn:frank1}
\theta\mathbf n+\sum_{j=1}^J \rho^{(j)}\mathbf b^{(j)}=0,
\end{flalign}
or
\begin{flalign}\label{eqn:frank2}
\theta\mathbf n+\sum_{j=1}^J \frac{\varrho^{(j)}\mathbf b^{(j)}}{\sqrt{R^2+R'^2}}=0.
\end{flalign}
An equivalent form is
 \begin{flalign}\label{eqn:frank3}
\left(\theta\sqrt{R^2+R'^2}\right)\mathbf n+\sum_{j=1}^J \varrho^{(j)}\mathbf b^{(j)}=0.
\end{flalign}

Using the Frank's formula in Eq.~\eqref{eqn:frank1} or \eqref{eqn:frank2},   the misorientation angle $\theta$ at each point on the grain boundary can be calculated by
\begin{flalign}\label{theta_calculation1}
\theta=-\sum_{j=1}^J \rho^{(j)}\left(\mathbf b^{(j)}\cdot\mathbf n\right)
=-\sum_{j=1}^J \frac{\varrho^{(j)}\left(\mathbf b^{(j)}\cdot\mathbf n\right)}{\sqrt{R^2+R'^2}}.
\end{flalign}
Note that an alternative formula to calculate the misorientation angle is
$\theta= \|\sum_{j=1}^J \rho^{(j)}\mathbf b^{(j)}\|=\|\sum_{j=1}^J \varrho^{(j)}\mathbf b^{(j)}\|/\sqrt{R^2+R'^2}$.
Integrating Eq.~\eqref{theta_calculation1} over the grain boundary, we have
 \begin{flalign}\label{eqn:theta_noncircular}
\theta =-\frac{1}{L}\int_0^{2\pi}\sum_{j=1}^J \varrho^{(j)}\left(\mathbf b^{(j)}\cdot{\mathbf n}\right)d\alpha,
\end{flalign}
where $L$ is the circumference of the grain boundary. It is more convenient to use this formula to evaluate $\theta$ during the evolution of the grain boundary in which the shape of the grain boundary and the dislocation densities on it change.

\subsection{Frank's formula maintained by the long-range dislocation interaction}\label{sec:Frank-longrange}

It has been shown that the Frank's formula is equivalent to the cancellation of the long-range elastic fields  generated by the constituent dislocations of the grain boundary \citep{Frank,Bilby,Zhu2014175}. In the continuum model in Eqs.~\eqref{eqn:vn} and \eqref{eqn:vp} (or Eqs.~\eqref{eqn:vr} and \eqref{eqn:vpr}), the Frank's formula is maintained by the continuum long-range dislocation interaction force. This is because the continuum long-range dislocation force is much stronger than the local force  except for very small size grain (comparable with the inter-dislocation distance on the grain boundary) \citep{Zhu2014175}. Starting from an equilibrium grain boundary with vanishing continuum long-range interaction, approximately during the motion of the grain boundary, the continuum long-range  interaction will remain vanishing and the Frank's formula will always hold  (except when the grain size is very small), otherwise the long-range interaction would lead to an increase in the total energy.

A proof of the equivalence of the Frank's formula and cancellation of the  long-range elastic fields  generated by the constituent dislocations for a curved grain boundary can be found in \citet{Zhu2014175}. Here we give an alternative calculation in two dimensions to show their equivalence, see \ref{sec:equivalenceproof}.

\section{Grain rotation and coupling and sliding motions}\label{sec:compare}
In this section, we present formulas for the rate of change of the misorientation angle $d\theta/dt$ (grain rotation) and the tangential velocities of the coupling and sliding motions of the grain boundary. Derivation of these formulas are based on the Frank's formula, which is maintained by the long-range dislocation interaction in our continuum model in Eqs.~\eqref{eqn:vn} and \eqref{eqn:vp} (or Eqs.~\eqref{eqn:vr} and \eqref{eqn:vpr}).

\subsection{Grain rotation and coupling and sliding velocities}\label{sec:rotation}
Using the Frank's formula in Eq.~\eqref{eqn:frank2}, it can be calculated that the rate of change of the misorientation angle $\theta$ is
\begin{flalign}\label{eqn:dtheta_dt}
\frac{d\theta}{dt}
=-\frac{\theta}{R}\frac{\partial R}{\partial t}
+\frac{1}{R}\sum_{j=1}^J \left(\mathbf b^{(j)}\cdot\hat{\mathbf R}\right) \frac{\partial \varrho^{(j)}}{\partial t},
\end{flalign}
where $\hat{\mathbf R}=\mathbf R/R$ is the outward radial direction of the grain boundary. The time derivatives $\partial R/\partial t$ and $\partial \varrho^{(j)}/\partial t$ in this grain rotation formula are given by our continuum model in Eqs.~\eqref{eqn:vr} and \eqref{eqn:vpr}.
Derivation of this
grain rotation formula from Frank's formula is given in \ref{eqn:grainrotationderivation}. It is interesting to see that the grain rotation formula in Eq.~\eqref{eqn:dtheta_dt} depends on the net Burgers vector in the radial direction, whereas the formulation of the misorientation angle  in Eq.~\eqref{eqn:theta_noncircular} depends on the net Burgers vector in the normal direction of the grain boundary.

The tangential velocity of the grain boundary is calculated based on the change of misorientation angle and the fixed center, which is
$v_{\|}=R\cos\lambda\frac{d\theta}{dt}=\frac{R^2}{\sqrt{R^2+R'^2}}\frac{d\theta}{dt}$,
where $\lambda$ is recalled to be the angle between the normal direction of the grain boundary and its inward radial direction, and $\cos\lambda=R/\sqrt{R^2+R'^2}$, see Fig.~\ref{fig:geometry}. Using the grain rotation formula in Eq.~\eqref{eqn:dtheta_dt}, the tangential velocity can be written as
\begin{flalign}\label{eqn:coupling0}
v_{\|}=\theta v_n+\frac{R}{\sqrt{R^2+R'^2}}\sum_{j=1}^J \left(\mathbf b^{(j)}\cdot\hat{\mathbf R}\right) \frac{\partial \varrho^{(j)}}{\partial t}.
\end{flalign}
Here we have used $v_n=-(\partial R/\partial t)\cos\lambda$.

The grain rotation formula in Eq.~\eqref{eqn:dtheta_dt} and the tangential velocity formula in Eq.~\eqref{eqn:coupling0} show that the grain rotation (or the tangential velocity) can be decomposed into the rotation (or the tangential velocity) due to the normal motion of the grain boundary with dislocation conservation (the first term in Eq.~\eqref{eqn:dtheta_dt} or \eqref{eqn:coupling0}) and that due to the evolution of the constituent dislocations on the grain boundary (the second term in Eq.~\eqref{eqn:dtheta_dt} or \eqref{eqn:coupling0}). The former is recalled to be the coupling motion of the grain boundary and is purely geometric \citep{Cahn20021,Cahn20044887}.
The latter includes the sliding motion of the grain boundary which is recalled to be the rigid-body translation  independent of the normal motion of the grain boundary \citep{Shewmon1966}.
The grain boundary dislocations are not conserved during the sliding motion \citep{Cahn20044887,Trautt20122407}.
This can be understood as follows. When the misorientation angle $\theta$ changes, say decreases, and the grain boundary profile remains the same, the inter-dislocation distance increases, which means the total number of the grain boundary dislocations decreases. This can also be seen from our formulation that the change of dislocation densities $\partial \varrho^{(j)}/\partial t$ in the second term in the grain rotation formula in Eq.~\eqref{eqn:dtheta_dt} or the tangential velocity formula in Eq.~\eqref{eqn:coupling0} includes the motion of existing dislocations on the grain boundary (the first term on the right-hand side of Eq.~\eqref{eqn:vpr}) and the changes of dislocation densities due to dislocation reactions (the second and third terms on the right-hand side of Eq.~\eqref{eqn:vpr}). The effect of the motion of existing dislocations on the grain boundary is to maintain the Frank's formula along with the motion of the grain boundary and reactions of dislocations.

The integral form the the grain rotation formulation is:
 \begin{flalign}\label{eqn:theta_t_noncircular}
\frac{d\theta}{dt} =-\frac{\theta}{L}\frac{d L}{d t}-\frac{1}{L}\int_0^{2\pi}\sum_{j=1}^J \frac{\partial \varrho^{(j)}}{\partial t}\left(\mathbf b^{(j)}\cdot{\mathbf n}\right)d\alpha.
\end{flalign}
Using Eq.~\eqref{eqn:vpr}, it can be written as
 \begin{flalign}\label{eqn:theta_t_noncircular1}
\frac{d\theta}{dt} =&-\frac{\theta}{L}\frac{d L}{d t}
+\frac{M_{\rm d} }{L}\int_0^{2\pi}\kappa\sum_{j=1}^J  \varrho^{(j)} \left({\mathbf f}_{\rm long}^{(j)}\cdot \mathbf T\right)\left(\mathbf b^{(j)}\cdot{\mathbf T}\right)d\alpha\nonumber\\
&+\frac{ M_{\rm t}}{L}\int_0^{2\pi}\sum_{j=1}^J\frac{\partial \gamma}{\partial \varrho^{(j)}}
\left(\mathbf b^{(j)}\cdot{\mathbf n}\right)d\alpha
+\frac{M_{\rm a}}{L}\int_0^{2\pi}\sum_{j=1}^J \left( \frac{\varrho^{(j)}}{|\varrho^{(j)}|} \ {\mathbf f}_{\rm app}^{(j)}\cdot\mathbf T\right)'\left(\mathbf b^{(j)}\cdot{\mathbf n}\right)d\alpha.
\end{flalign}
Here we have used the equation $\sum_{j=1}^J  \varrho^{(j)} (\mathbf b^{(j)}\cdot{\mathbf T})=0$ by using Eq.~\eqref{eqn:frank3}.

\subsection{Comparisons with the models of Cahn-Taylor and Shewmon}

 Our grain rotation and tangential velocity formulas in Eqs.~\eqref{eqn:dtheta_dt} and \eqref{eqn:coupling0} are derived based on the motion and reaction of the grain boundary dislocations and Frank's formula.
They can be considered as justifications in the case of low angle grain boundaries for the
 unified approach to the grain boundary coupling and sliding motions of \citet{Cahn20044887},  in which existence of the coupling effect and superposition of the coupling and sliding motions are assumptions.
Our continuum model also generalizes the Cahn-Taylor model
 by  incorporating detailed formulas of the driving forces for the  grain boundary  coupling and sliding tangential motions that depend on the constituent dislocations, their Burgers vectors, and the grain boundary shape,
 as well as the  shape change of the grain boundaries.
 For the special case of a circular grain boundary with a single Burgers vector, our   tangential velocity and grain rotation formulas reduce to the equations in the Cahn-Taylor model as illustrated below.

When the grain boundary is circular,
$R(\alpha)$ is constant, $R'(\alpha)=0$, $\kappa=1/R$, and $\mathbf n=-\hat{\mathbf R}$. At a point on the circular grain boundary where all the dislocations have the same Burgers vector $\mathbf b$ that is in the direction $\hat{\mathbf R}$, i.e., $b=\mathbf b\cdot \hat{\mathbf R}$,  our grain rotation and tangential velocity formulas in Eqs.~\eqref{eqn:dtheta_dt} and \eqref{eqn:coupling0} become
\begin{flalign}
\frac{d\theta}{dt}&=-\frac{\theta}{R} \frac{dR}{dt}-\frac{M_{\rm t}b}{R} \label{eqn:circularsinglerot} \frac{\partial\gamma}{\partial{\varrho}}+\frac{M_{\rm a}b^2 }{R}\sigma_{\rm rt},\\
v_{\|}&=\theta v_n-M_{\rm t} b\frac{\partial\gamma}{\partial{\varrho}}+M_{\rm a}b^2 \sigma_{\rm rt},\label{eqn:circularsingle}
\end{flalign}
where $\sigma_{\rm rt}=\hat{\mathbf R}\cdot \pmb\sigma_{\rm app}\cdot \mathbf T$ is the shear component of the applied stress along the grain boundary.

For this special case of circular grain boundary with single Burgers vector, the  tangential velocity of the grain boundary given previously by the Cahn-Taylor model \citep{Cahn20044887,Trautt20122407} is
\begin{flalign}
\frac{d\theta}{dt}&=-\frac{\beta}{R} \frac{dR}{dt} + \frac{S}{R}\left(\sigma-\frac{\gamma'(\theta)}{R}\right),\label{mishin_eqn13rot}\\
v_{\|}&=\beta v_n + S\left(\sigma-\frac{\gamma'(\theta)}{R}\right),\label{mishin_eqn13}
\end{flalign}
where $\beta=\theta$ is the coupling factor, $\gamma'(\theta)=d\gamma/d\theta$,  $S$ is the sliding coefficient, and $\sigma$ is the applied shear stress.
 Comparing Eqs.~\eqref{eqn:circularsingle} and \eqref{mishin_eqn13} (same for the comparison between Eqs.~\eqref{eqn:circularsinglerot} and \eqref{mishin_eqn13rot}),
we can see that our formula gives the same coupling term (the first term in Eq.~\eqref{eqn:circularsingle}) as that in the Cahn-Taylor model.
The last two terms in our formula in Eq.~\eqref{eqn:circularsingle} that accounts for  the grain boundary sliding are also the same as
the corresponding terms in the Cahn-Taylor model in Eq.~\eqref{mishin_eqn13} based on the relations $\theta=b/D=\varrho b/R$ and
$\partial\gamma/\partial \varrho=\gamma'(\theta)b/R$ if we further assume $\varrho>0$ and $S=M_{\rm t}b^2=M_{\rm a}b^2$.

Note that at an arbitrary point on the circular low angle grain boundary, dislocation densities of multiple Burgers vectors are in general nonzero, and our full formulas in Eqs.~\eqref{eqn:dtheta_dt} and \eqref{eqn:coupling0} should be used  instead of the above simple formulas. Moreover, the circular grain boundary may change its shape during its evolution with dislocation reactions,  see the simulations in Sec.~\ref{sec:simulation}.

 We also remark that in the generalized  Cahn-Taylor theory   based on the assumptions of coupling effect and mass transfer via surface diffusion along the grain boundary \citep{Taylor2007493}, alternative formulations for grain rotation and grain boundary tangential velocities due to the coupling and sliding as well as the shape-preserving motion for a non-circular grain boundary have been derived.

 In the model of \citet{Shewmon1966} for grain rotation by sliding, the rate of grain rotation is proportional to $\gamma'(\theta)$. This is included in the unified model by \citet{Cahn20044887} as shown in Eq.~\eqref{mishin_eqn13rot}. In our grain rotation formula in Eq.~\eqref{eqn:dtheta_dt}, the rate of rotation due to sliding is proportional to the change rates of the grain boundary dislocation densities, which together with the Frank's formula (maintained by the long-range elastic interaction) lead to the change of misorientation angle. Using densities of the constituent dislocations instead of the misorientation angle enables us to incorporate both the coupling and sliding motions of the grain boundary in the continuum model as discussed previously. In the simple case of a point on a circular grain boundary where all the dislocations have the same Burgers vector, our grain rotation formula includes the sliding-included rotation  proportional to $\partial\gamma/\partial \varrho=\gamma'(\theta)b/R$ as shown in Eq.~\eqref{eqn:circularsinglerot} and discussed above.

\subsection{Shape-preserving grain boundary motion under pure coupling}\label{sec:preserving}
Assume that the Frank's formula  holds during the evolution of the grain boundary. Consider the case when the tangential motion of the grain boundary is pure coupling,
 i.e. its constituent dislocations are conserved. From the simulations of discrete dislocation dynamics and our continuum model shown in Sec.~\ref{sec:simulation}, the constituent dislocations are moving in the inward radial direction, i.e. $\partial \varrho^{(j)}/\partial t=0$, $j=1,2,\cdots,J$. Under these conditions, the grain rotation formula in Eq.~\eqref{eqn:dtheta_dt} gives
 \begin{flalign}
\frac{d\theta}{dt}=-\frac{\theta}{R}\frac{\partial R}{\partial t}.
\end{flalign}
Integration of this equation gives
\begin{flalign}\label{eqn:Rfixshape}
R(\alpha,t)=\frac{\theta(0)}{\theta(t)}R(\alpha,0).
\end{flalign}
That is, the shape of the grain boundary does not change during the evolution. This was solved previously by
\citet{Taylor2007493} based on the assumption of the coupling effect and mass transfer via surface diffusion along the grain boundary.

In addition to this shape-preserving property, we also have the following identities during the grain boundary motion under conservation of dislocations by using Eqs.~\eqref{eqn:dtheta_dt} and \eqref{eqn:theta_t_noncircular}:
\begin{flalign}
\theta(t)R(\alpha,t)&=\theta(0)R(\alpha,0),\vspace{1ex}\label{eqn:Rtheta}\\
\theta(t)L(t)&=\theta(0)L(0).\label{eqn:Ltheta}
\end{flalign}
Recall that $L=\int_0^{2\pi}\sqrt{R^2+R'^2}d\alpha$ is the circumference of the grain boundary.

  Finally, we remark that  by Eq.~\eqref{eqn:theta_noncircular},
 \begin{flalign}\label{eqn:theta_noncircular1}
L\theta =-\int_0^{2\pi}\sum_{j=1}^J \varrho^{(j)}\left(\mathbf b^{(j)}\cdot{\mathbf n}\right)d\alpha,
\end{flalign}
which holds generally no matter whether the dislocation densities  $\varrho^{(j)}$, $j=1,2,\cdots,J$ change or not.

\section{Numerical Simulations}\label{sec:simulation}
In this section, we perform numerical simulations using our continuum model in Eqs.~\eqref{eqn:vr} and \eqref{eqn:vpr} to study the motion of grain boundaries and evolution of their dislocation structure in two dimensions.
 We focus on the grain boundaries in fcc crystals, in which there are $J=6$ Burgers vectors of the $<110>$ type with the same  magnitude $b$. The cross-section plane of the grain boundaries is the $(111)$ plane, and the directions $[\bar{1}10], [\bar{1}\bar{1}2]$ and $[111]$ are chosen to be the $x, y, z$ directions, respectively. In this coordinate system, the six Burgers vectors are
$\mathbf{b}^{(1)} = \left(1,0,0\right)b$, $\mathbf{b}^{(2)} = \left(\frac{1}{2},\frac{\sqrt{3}}{2},0\right)b$, $\mathbf{b}^{(3)} = \left(\frac{1}{2},-\frac{\sqrt{3}}{2},0\right)b$, $\mathbf{b}^{(4)} = \left(0,\frac{\sqrt{3}}{3},\frac{\sqrt{6}}{3}\right)b$, $\mathbf{b}^{(5)} = \left(\frac{1}{2},\frac{\sqrt{3}}{6},-\frac{\sqrt{6}}{3}\right)b$, and $\mathbf{b}^{(6)} = \left(-\frac{1}{2},\frac{\sqrt{3}}{6},-\frac{\sqrt{6}}{3}\right)b$.  The rotation axis is fixed as the $[111]$ direction. The misorientation angle of the initial grain boundary in the simulations is $5^\circ$. We choose the Poisson's ratio $\nu=0.347$, which is value of aluminum.

In our simulations, we neglect the applied stress and the bulk energy difference, that is, $\pmb \sigma_{\rm app}=\mathbf 0$ (accordingly, $\mathbf f^{(j)}_{\rm app}=0$, $j=1,2,\cdots,J$) and $p=0$
 in Eqs.~\eqref{eqn:vr} and \eqref{eqn:vpr}.
 The grain boundary motion by  Eq.~\eqref{eqn:vr} is solved using standard forward Euler scheme. The singular integrals in the long-range force $\mathbf f_{\rm long}^{(j)}$ in Eqs.~\eqref{eqn:flong0}--\eqref{eqn:flong1} are calculated using the trapezoidal rule with local approximation at the singular point \citep{SI1988}. The evolution of the dislocation structure in Eq.~\eqref{eqn:vpr} is solved by the following splitting scheme in time:
\begin{flalign}
\frac{\eta^{(j)}_*-\eta^{(j)}(t_n)}{\Delta t}&=-M_{\rm d} \frac{{\eta^{(j)}}'}{\sqrt{R^2+R'^2}}
\left[ {\mathbf f}_{\rm long}^{(j)}\cdot \mathbf T
+v_n\tan\lambda\right]_{t=t_n}, \label{eqn:vp11}\\
\frac{\varrho^{(j)}(t_{n+1})-\varrho^{(j)}_*}{\Delta t}&=\left.-M_{\rm t}\frac{\partial \gamma}{\partial \varrho^{(j)}}\right|_{t=t_n},
\end{flalign}
where $\Delta t$ is the time step. Here recall that $\varrho(\alpha)=\eta'(\alpha)$ or $\eta(\alpha)=\int^\alpha_0 \varrho(\alpha_1)d\alpha_1$. The spatial derivative ${\eta^{(j)}}'$  in Eq.~\eqref{eqn:vp11} is calculated using the upwind scheme, and a cutoff parameter is used to set ${\eta^{(j)}}'=0$ when the calculated value is very small.

We start from a grain boundary with the equilibrium dislocation structure that satisfies the Frank's formula and has the lowest energy, which can be calculated pointwisely using the method in \citet{Zhang2017}.

We also perform discrete dislocation dynamics simulations to validate our continuum model.
 The formulation of the interaction between dislocations can be found, e.g., in \citet{HL}. In fact,
  the Peach-Koehler force on a dislocation with Burgers vector $\mathbf b^{(i)}$ and line direction $\mathbf t^{(i)}$ is $(\pmb\sigma\cdot {\mathbf b^{(i)}}) \times \mathbf t^{(i)}$, where $\pmb \sigma$ is the stress. The stress at a point $(x,y)$ generated by a dislocation located at point $(x_1,y_1)$ with Burgers vector $\mathbf b^{(j)}=(b^{(j)}_1,b^{(j)}_2)$ and $+z$ line direction is $\pmb\sigma^{(j)}(x,y)=\mathbf G_1(x,y;x_1,y_1)b^{(j)}_1+\mathbf G_2(x,y;x_1,y_1)b^{(j)}_2$, where $\mathbf G_1$ and $\mathbf G_2$ are given in Eqs.~\eqref{G1}  and \eqref{G2}.

\subsection{Circular grain boundary}\label{sec:circle}
We firstly consider the motion of an initially circular grain boundary with radius $R_0=140b$.  The calculated  equilibrium dislocation structure \citep{Zhang2017} on this initial grain boundary is shown in Fig~\ref{circle_motion}a, which contains three arrays of dislocations with Burgers vectors $\mathbf{b}^{(1)}$, $\mathbf{b}^{(2)}$, and $\mathbf{b}^{(3)}$, respectively.  The distribution of these dislocations are represented by the dislocation density potential functions $\eta^{(1)}$, $\eta^{(2)}$, and ${\eta^{(3)}}$, respectively, see Fig~\ref{circle_motion}b.  Recall that the dislocation density potential functions are defined on the grain boundary such that the constituent dislocations are located at their integer-valued contour lines. Also recall that the dislocation density per unit polar angle $\varrho^{(j)}$ and the dislocation density per unit length $\rho^{(j)}$ can be calculated from $\eta^{(j)}$ using Eqs.~\eqref{eqn:density_angle} and \eqref{eqn:densityperlength}.
It can be seen that in this equilibrium dislocation structure,
each array of dislocations (with the same Burgers vector) concentrate in two portions of the grain boundary, and on each point on the grain boundary, there are nonzero dislocation densities of two different Burgers vectors (except for some extreme points).
In  Fig~\ref{circle_motion}b, the locations of dislocations using discrete dislocation model are also shown using the dislocation density potential functions.

\subsubsection{Motion under dislocation conservation}\label{sec:circle_conservation}

\begin{figure}[htbp]
\centering
 \includegraphics[width=0.65\linewidth]{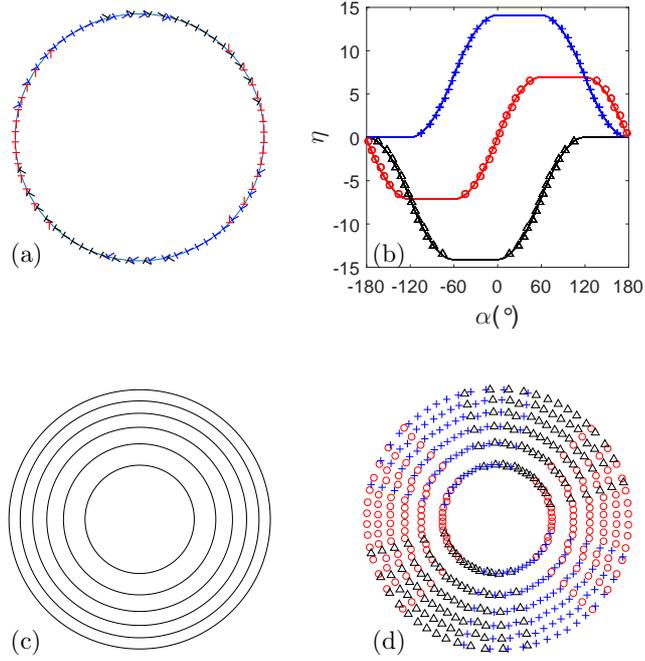}
     \caption{Grain boundary motion under dislocation conservation. The initial grain boundary is circular. (a) Equilibrium dislocation structure on the initial circular grain boundary. There are three arrays of dislocations with  Burgers vectors $\mathbf{b}^{(1)}$ (red), $\mathbf{b}^{(2)}$ (black), and $\mathbf{b}^{(3)}$ (blue), respectively. (b) Evolution of dislocation density potential functions of these three arrays of dislocations on the grain boundary (same colors as in (a)). The curves show the results of the continuum model and the dots show the results using the discrete dislocation model. (c) Motion of the grain boundary (shrinkage) by using our continuum model.  (d)  Motion of the grain boundary by using the discrete dislocation model. In (c) and (d),  the grain boundary is plotted at uniform time intervals starting with the outer most one. }
    \label{circle_motion}
\end{figure}

In the continuum model in Eqs.~\eqref{eqn:vr} and ~\eqref{eqn:vpr} (or Eqs.~\eqref{eqn:vn} and ~\eqref{eqn:vp}), recall that the mobility  $M_{\rm d}$ is associated with the conservative law motion of dislocations and the mobility $M_{\rm t}$ is associated with  dislocation reactions. Recall that the dislocation structure on the grain boundary is stable and energy barriers have to be overcome for dislocations to react. In the simulations in this subsection, we set $M_{\rm t}=0$, i.e., the dislocations move entirely by the conservation law without dislocation reaction. This can happen to low angle grain boundaries whose sizes are not very small, as observed in atomistic \citep{Cahn20021} and phase-field crystal \citep{Wu2012407} simulations.

Figure~\ref{circle_motion}c and \ref{circle_motion}d show the evolution  of the grain boundary obtained by using our continuum model and the discrete discrete model, respectively, under the same initial configuration and dislocation mobility.  In the simulation results of the continuum model
shown in Fig~\ref{circle_motion}c, the circular grain boundary shrinks with increasing rate and keeps the circular shape during the shrinkage. This evolution is in excellent agreement with the simulation results using the discrete dislocation model shown in  Fig~\ref{circle_motion}d.

The evolutions of the dislocation structure during the shrinking of this grain boundary by using the continuum model and the discrete dislocation model are shown in Fig~\ref{circle_motion}b, based on the dislocation density potential functions $\eta^{(1)}$, $\eta^{(2)}$, and ${\eta^{(3)}}$ for dislocations with Burgers vectors $\mathbf{b}^{(1)}$, $\mathbf{b}^{(2)}$, and $\mathbf{b}^{(3)}$, respectively, as illustrated at the beginning of this section. We can see from this figure that the dislocation density potential functions using both the continuum model and the discrete model  maintain their initial profiles during the evolution, meaning that the dislocation densities per unit polar angle do not change. This shows that all the dislocations move in the inward radial direction of the grain boundary using both models, as can also be seen from the results of the discrete model shown in Fig~\ref{circle_motion}d. The Frank's formula always holds and the long-range dislocation interaction vanishes during the motion of this circular grain boundary.

\begin{figure}[htbp]
\centering
      \includegraphics[width=0.9\linewidth]{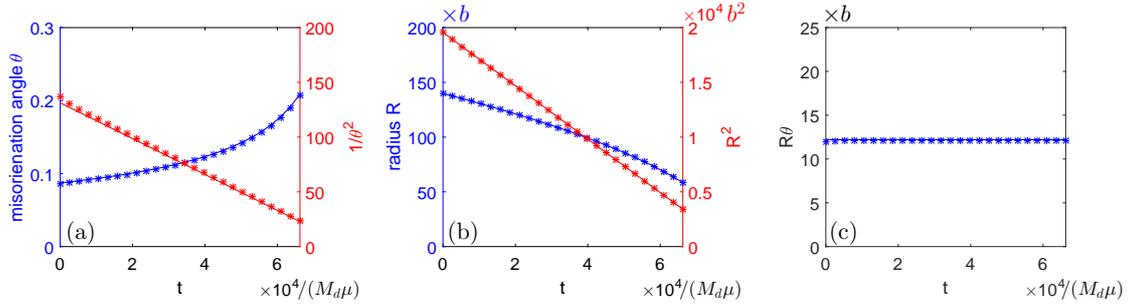}
    \caption{Grain boundary motion under dislocation conservation. The initial grain boundary is circular. (a) The misorientation angle $\theta$ as a function of evolution time. (b) The radius $R$ as a function of evolution time. (c) $R\theta$ keeps  constant during the evolution. The curves show the results of the continuum model and the dots show the results using the discrete dislocation model.}
    \label{circle_R_rotation}
\end{figure}

In this simulation, the grain rotates by pure coupling effect as it shrinks.
We calculate the misorientation angle $\theta$ during the shrinkage of the circular grain boundary by Eq.~\eqref{theta_calculation1} (where $R'\equiv0$).
 Fig.~\ref{circle_R_rotation}a shows the misorientation angle $\theta$ and $1/{\theta}^2$ as functions of the evolution time $t$. In Fig.~\ref{circle_R_rotation}b, the radius $R$ of the evolving circular grain boundary, as well as the quantity $R^2$ are shown in terms of $t$.
 It can be seen that the misorientation angle $\theta$ increases as the grain boundary shrinks,
 and the changing rates of both $\theta$ and $R$   are increasing. More precisely, the quantities $1/{\theta}^2$ and $R^2$ decrease linearly with $t$.
  We also examine the relation between the radius $R$ and misorientation $\theta$, and it can be seen that their multiplication $R\theta$ is constant during the evolution as shown in Fig.~\ref{circle_R_rotation}c. This relation agrees with the theoretical prediction by Eq.~\eqref{eqn:Rtheta}.
  Moreover, by the grain boundary evolution equation in  Eq.~\eqref{eqn:vr}, for this circular grain boundary,  the long-range force ${\mathbf f}_{\rm long}$ vanishes as discussed above, and the evolution equation is reduced to $v_n=\frac{M_{\rm d}\mu b^2}{4\pi(1-\nu)R}$
    and $\frac{dR}{dt}=-v_n$.
  This implies that $R(t)^2$ is a linear function of $t$: $R(t)^2=R(0)^2-\frac{M_{\rm d}\mu b^2}{2\pi(1-\nu)}t$.
 Accordingly, $1/\theta(t)^2$ is also a linear function of $t$ due to the fact that $R(t)\theta(t)$ remains constant.   The simulation results discussed above agree with these theoretical predictions. Especially, both the slope of the numerical $R^2(t)$ curve in Fig.~\ref{circle_R_rotation}b  and the theoretical formula of the slope $\frac{d R^2}{dt}=-\frac{M_{\rm d}\mu b^2}{2\pi(1-\nu)}$ give the same value $-0.2437M_{\rm d}\mu b^2$.

\begin{figure}[htpb]
\centering
    \includegraphics[width=0.9\linewidth]{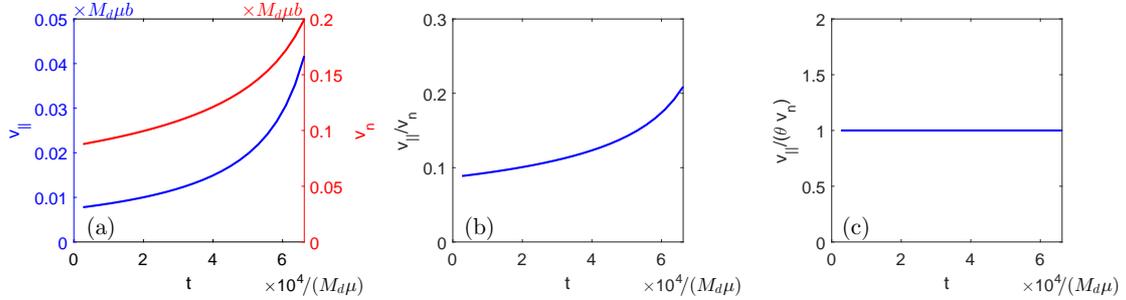}
     \caption{Grain boundary motion under dislocation conservation. The initial grain boundary is circular. (a) The tangential velocity $v_{\|}$ and normal velocity $v_{n}$ of the grain boundary. (b) The ratio  $v_{\|}/v_{n}$. (c) Verification of the coupling relation $v_{\|}=\theta v_n$.}
    \label{circle_v}
\end{figure}

 During the shrinkage of this circular grain boundary, the tangential velocity $v_{\|}$ and the normal velocity $v_n$ of the grain boundary should satisfy the coupling relation
$v_{\|}=\theta v_n$
 by our Eq.~\eqref{eqn:coupling0} (with $\partial \varrho^{(j)}/\partial t=0$) or the theory of \citet{Cahn20044887}. We examine this relation by our continuum simulation as shown in Fig.~\ref{circle_v}.
The tangential velocity $v_{\|}$ and the normal velocity $v_n$ during the evolution of the circular grain boundary are plotted in Fig.~\ref{circle_v}a, where   $v_{\|}$ is calculated by
$v_{\|}=R d\theta/dt$.
 The relation of the two velocities obtained from our simulation results are shown in Fig.~\ref{circle_v}b and Fig.~\ref{circle_v}c.  Excellent agreement can be seen between the numerical and theoretical values for the coupling relation $v_{\|}=\theta v_n$.

These simulation results of the continuum model and the discrete dislocation model agree excellently with the available atomistic \citep{Cahn20021,Trautt20122407} and phase-field crystal \citep{Wu2012407} simulations, and the Cahn-Taylor theory \citep{Cahn20044887} which is a mechanism  without details of the dislocation structure on the grain boundary.

\subsubsection{With dislocation reaction}\label{sec:circle_reaction}

\begin{figure}[htbp]
\centering
    \includegraphics[width=0.6\linewidth]{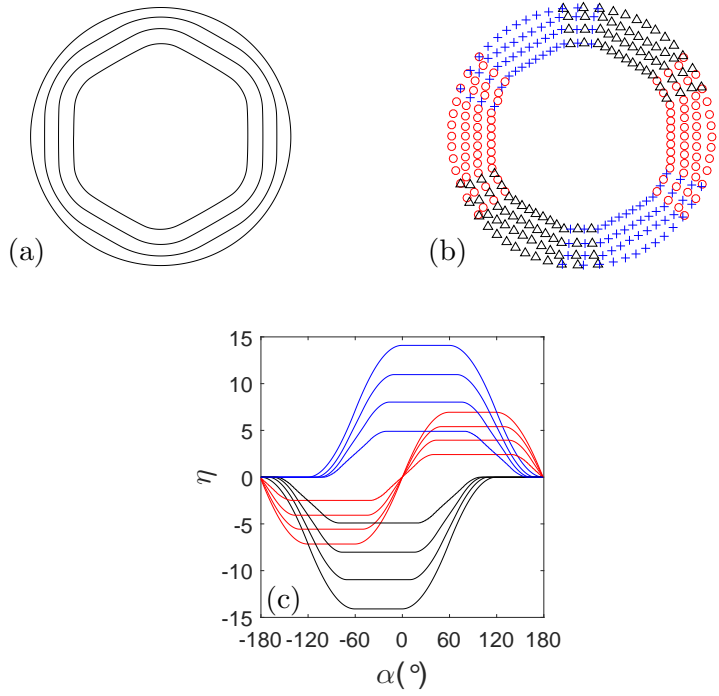}
    \caption{Grain boundary motion with dislocation reaction: $M_{\rm t}b/ M_{\rm d}=0.286\times 10^{-4}$. The initial grain boundary is circular. (a) Grain boundary motion using our continuum model. The grain boundary is plotted at uniform time intervals starting with the outer most one.
    (b) Grain boundary motion by using discrete dislocation dynamics model, where two pairs of dislocations in each of the three arrays of dislocations with different Burgers vectors are removed in the initial dislocation structure of the circular grain boundary. (c) Evolution of the dislocation density potential functions $\eta^{(1)}$, $\eta^{(2)}$, and ${\eta^{(3)}}$ of the three arrays of dislocations with  Burgers vectors $\mathbf{b}^{(1)}$ (red), $\mathbf{b}^{(2)}$ (black), and $\mathbf{b}^{(3)}$ (blue), respectively.
    }
    \label{circle_motion1}
\end{figure}

In this subsection, we perform simulations using our continuum model for the case in which dislocation reaction is not negligible during the motion of the grain boundary, i.e., $M_{\rm t}\neq 0$. Recall again that the dislocation structure on the grain boundary is stable and energy barriers have to be overcome for dislocations to react.
The simulation results from the initial, circular grain boundary with $M_{\rm t}b/ M_{\rm d}=0.286\times 10^{-4}$ are shown in Fig.~\ref{circle_motion1}a. The initial circular grain boundary gradually changes to hexagonal shape as it shrinks.
Evolution of the dislocation density potential functions $\eta^{(j)}$, $j=1,2,3$, is shown in Fig.~\ref{circle_motion1}c. The amplitude of each $\eta^{(j)}$ is decreasing,  meaning that the dislocations react and the number of dislocations of each Burgers vector is reduced.
For the purpose of comparison, in the discrete dislocation dynamics simulation, we manually remove two pairs of dislocations in each of the three arrays of dislocations with different Burgers vectors in the initial dislocation structure of the circular grain boundary,
 and the simulation results are shown in Fig.~\ref{circle_motion1}b. The grain boundary profile obtained in this discrete dislocation dynamics simulation also shows that the grain boundary evolves towards a hexagonal shape. Excellent qualitative agreement can be seen from the simulation results using these two models. Note that in the simulations using our continuum model, dislocations are removed continuously during the motion of the grain boundary. The reason to manually remove dislocations in the discrete dislocation dynamics simulations is that in order for the grain boundary dislocations to react, some energy barriers have to be overcome and these events are not included in the standard rules of discrete dislocation dynamics. In fact, the mechanisms of grain boundary dislocation reactions are non-trivial, e.g. by annihilation of dislocation pairs with opposite Burgers vectors located on opposite ends of the grain \citep{Cahn20021}, or by a chain of dissociation and association of the GB dislocations \citep{Trautt20122407}.

The evolution of this initially circular grain boundary into a hexagonal shape with dislocation reactions can be understood as follows. For a dislocation density $\varrho>0$ (same when $\varrho<0$), Eq.~\eqref{eqn:gb_density} gives $\frac{\partial \gamma}{\partial \varrho}=\frac{\mu b^2}{4\pi(1-\nu)\sqrt{R^2+R'^2}}\log\frac{\sqrt{R^2+R'^2}}{r_g\varrho}$ (neglecting the numerical cutoff parameter $\epsilon$).   Thus by Eq.~\eqref{eqn:vpr} with constant $M_{\rm t}\neq 0$, the rate of decrease of $\varrho$ due to $\frac{\partial \gamma}{\partial \varrho}$ is the largest when the density $\varrho$ is approaching $0$. It can be seen from Fig.~\ref{circle_motion1}c (or Fig.~\ref{circle_motion}a) that on the initial, circular grain boundary, dislocations with each Burgers vector occupy two arcs of the grain boundary, and these arcs of dislocations with different Burgers vectors overlap. During the evolution with $\frac{\partial \gamma}{\partial \varrho}$, all the arcs of dislocations shorten as discussed above and soon become disjoint. At this stage,  all the dislocations on each arc have the same Burgers vector and the arc becomes planar (symmetric tilt) under the long-range elastic interaction, leading to a hexagonal shape of the grain boundary.

\begin{figure}[htbp]
\centering
    \includegraphics[width=0.9\linewidth]{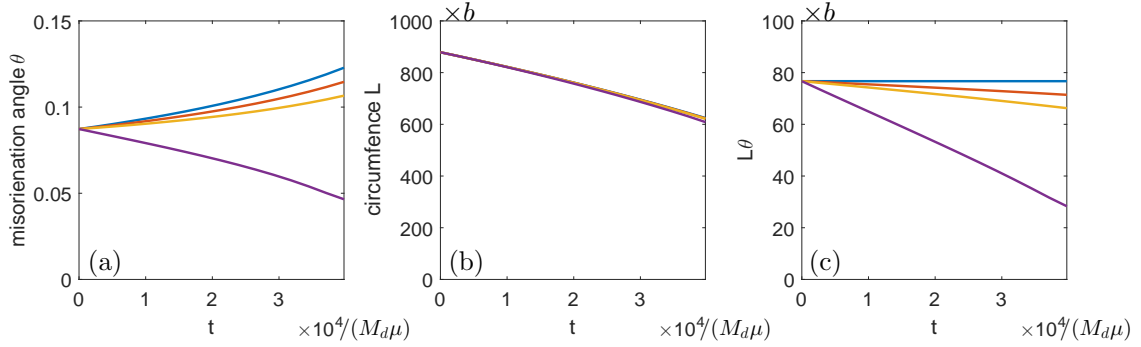}
    \caption{Grain boundary motion with dislocation reaction for different reaction mobility $M_{\rm t}$:  $M_{\rm t}b/ M_{\rm d}=0, 2.86\times 10^{-5}, 5.762\times 10^{-5}$, and $2.86\times 10^{-4}$ (from the top to the bottom). The initial grain boundary is circular. (a) Evolution of the misorientation angle $\theta$. (b) Evolution of the circumference  of the grain boundary $L$. (c) Evolution of $L\theta$. }
    \label{circle_theta_R_flocal}
\end{figure}

Evolution of the misorietation angle $\theta$ of this initial circular grain boundary with different values of reaction mobility $M_{\rm t}$ is shown in Fig.~\ref{circle_theta_R_flocal}a, using the formula in Eq.~\eqref{eqn:theta_noncircular}. In this case, the evolution of the misorientation angle $\theta$ is controlled by both the coupling effect that depends on the dislocation mobility $M_{\rm d}$ and the sliding effect that depends mainly on the mobility $M_{\rm t}$ due to dislocation reaction.
The coupling effect increases the misorientation angle $\theta$ while the sliding effect decreases $\theta$.
This can be understood as follows. At a point on the initial circular grain boundary where all the dislocations have the same Burgers vector $\mathbf b$ that is in the outward radial direction $\hat{\mathbf R}$,
by Eqs.~\eqref{eqn:dtheta_dt}, \eqref{eqn:vpr}, and \eqref{eqn:gb_density} together with the cancellation of the long-range force in the initial dislocation structure, we have
\begin{equation}
\frac{d\theta}{dt}=\frac{M_{\rm d}\mu b^2\theta}{4\pi(1-\nu)R^2}
-\frac{M_{\rm t}\mu b^3}{4\pi(1-\nu)R^2}\log \left(\frac{1}{e r_g\sqrt{\varrho^2/R^2  +\epsilon}}\right).
\end{equation}
The first term in this equation is due to the coupling effect and increases the misorientation angle $\theta$, whereas the second term is due to the sliding effect and decreases $\theta$.
Note that the above equation is just for qualitative understanding because here we have dislocations with three different Burgers vectors and the grain boundary shape also changes during the evolution. As can be seen from the simulation results in Fig.~\ref{circle_theta_R_flocal}a, when the dislocation reaction mobility $M_{\rm t}$ increases, meaning the sliding effect due to dislocation reaction is becoming stronger, the increase rate of $\theta$ is decreasing during the motion of the grain boundary, and when the sliding effect is strong enough, the misorientation angle $\theta$ is decreasing.

Evolutions of the circumference of the grain boundary $L$ and the product $L\theta$ are shown in Fig.~\ref{circle_theta_R_flocal}b and Fig.~\ref{circle_theta_R_flocal}c. Note that here we consider the circumference for a non-circular grain boundary instead of the radius $R$ for a circular one.
Recall that in the case without dislocation reaction discussed previously, $L\theta=2\pi R\theta$ keeps constant during the evolution of the grain boundary.
When sliding effect is not negligible, $L\theta$ is decreasing during the evolution of the grain boundary due to the fact that the sliding effect decreases the misorientation angle, as shown in Fig.~\ref{circle_theta_R_flocal}c, which is also in agreement with  Eq.~\eqref{eqn:theta_noncircular1}.
However, the evolution of the circumference of the grain boundary $L$ does not change much for these values of $M_{\rm t}b/M_{\rm d}$ compared with the case of $M_{\rm t}=0$, see Fig.~\ref{circle_theta_R_flocal}b.

\subsection{Elliptic grain boundary}\label{sec:ellipse}

We then consider the motion of a grain boundary with initial shape of ellipse. The long axis of the ellipse is $140b$ and the ratio of the  long axis to the short axis  is $1.5$, see the outer most curve in  Fig~\ref{ellipse_motion}a.  As in the circular grain boundary case, the equilibrium dislocation configuration on this elliptic grain boundary also consists of dislocations with Burgers vectors $\mathbf{b}^{(1)}$, $\mathbf{b}^{(2)}$, and $\mathbf{b}^{(3)}$, represented by dislocation density potential functions $\eta^{(1)}$, $\eta^{(2)}$, and ${\eta^{(3)}}$, respectively, as shown in Fig~\ref{ellipse_motion}c.
 The locations of dislocations on this grain boundary using discrete dislocation model are also shown by
 the outer most curve in  Fig~\ref{ellipse_motion}b and in
 Fig~\ref{ellipse_motion}c based on the dislocation density potential functions.

\subsubsection{Motion under dislocation conservation}\label{sec:ellipse_conservation}

\begin{figure}[htbp]
\centering
    \includegraphics[width=0.6\linewidth]{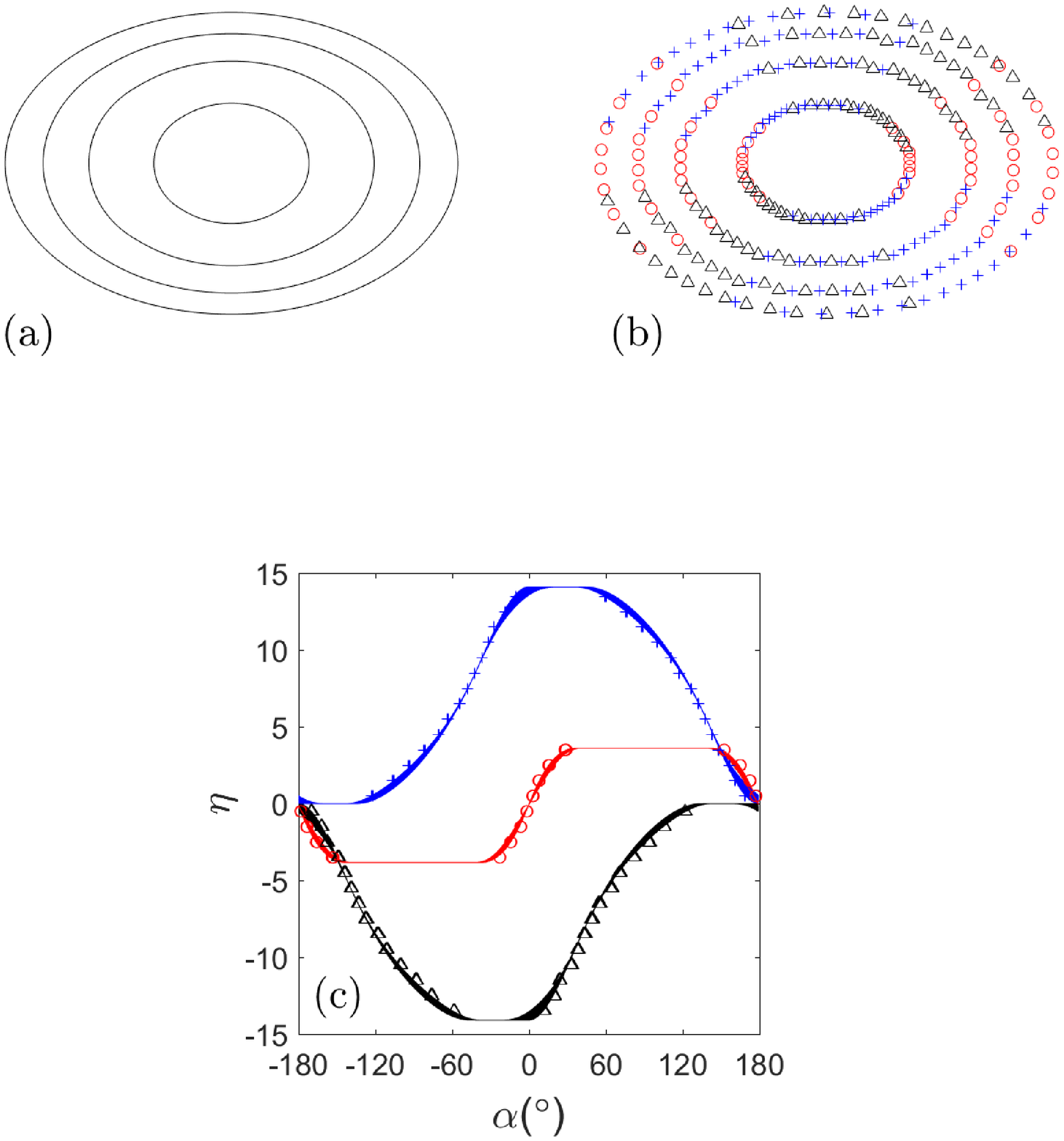}
    \caption{Grain boundary motion under dislocation conservation. The initial grain boundary is an ellipse.  (a) Motion of the grain boundary (shrinkage) by using our continuum model.  (b)  Motion of the grain boundary by using the discrete dislocation model. In (a) and (b),  the grain boundary is plotted at uniform time intervals starting with the outer most one. (c) Evolution of dislocation density potential functions $\eta^{(1)}$, $\eta^{(2)}$, and ${\eta^{(3)}}$,  which represent  dislocations on the grain boundary with  Burgers vectors $\mathbf{b}^{(1)}$ (red), $\mathbf{b}^{(2)}$ (black), and $\mathbf{b}^{(3)}$ (blue), respectively.
     The curves show the results of the continuum model and the dots show the results using the discrete dislocation model.}
    \label{ellipse_motion}
\end{figure}

We first  consider the motion of this initially elliptical grain boundary under dislocation conservation without dislocation reaction, i.e., $M_{\rm t}=0$. In this case, the grain boundary motion is pure coupling and there is no sliding effect. Simulation result  using our continuum model is shown in Fig~\ref{ellipse_motion}a. We can see that the elliptic grain boundary shrinks with increasing rate and its shape does not change during the evolution. We also perform simulation using the discrete dislocation dynamics model to validate our continuum model, and the result is plotted in Fig~\ref{ellipse_motion}b, which shows that all the dislocations move in the inward radial direction and the shape of the grain boundary is also preserved.
This is in agreement with the theoretical prediction in Sec.~\ref{sec:preserving} and
the prediction of the generalization of the Cahn-Taylor theory to noncircular grains based on assumptions of the coupling effect and mass transfer by surface diffusion \citep{Taylor2007493}.

Evolutions of the dislocation structure on this grain boundary represented by dislocation density potential functions $\eta^{(1)}$, $\eta^{(2)}$, and ${\eta^{(3)}}$ are shown in Fig~\ref{ellipse_motion}c. It can be seen that these functions do not change as the grain boundary shrinks, by using both  the discrete dislocation model and the continuum model up to numerical errors.  These simulation results indicate that the dislocation densities per unit polar angle, $\varrho^{(j)}={\eta^{(j)}}'$, $j=1,2,3$, do not change during the evolution of the grain boundary.
Since such dislocation structure on the evolving grain boundary always satisfies the Frank's formula,
the simulation results using both the discrete and continuum models justify the assumption that the Frank's formula holds during the evolution of the grain boundary, based on which the formulas and properties of the misorientation angle, grain rotation and coupling and sliding motions are derived  in Secs.~\ref{sec:Frank} and \ref{sec:compare}.

We believe the small errors in the numerical results of the continuum model compared with the discrete model seen in Fig~\ref{ellipse_motion}c are mainly due to the forward Euler type numerical schemes used in our simulations, see the beginning of this section.  Note that this paper focus on derivation of the continuum model. Efficient and accurate numerical schemes for this continuum model will be explored in the future work.

\begin{figure}[thbp]
\centering
    \includegraphics[width=0.9\linewidth]{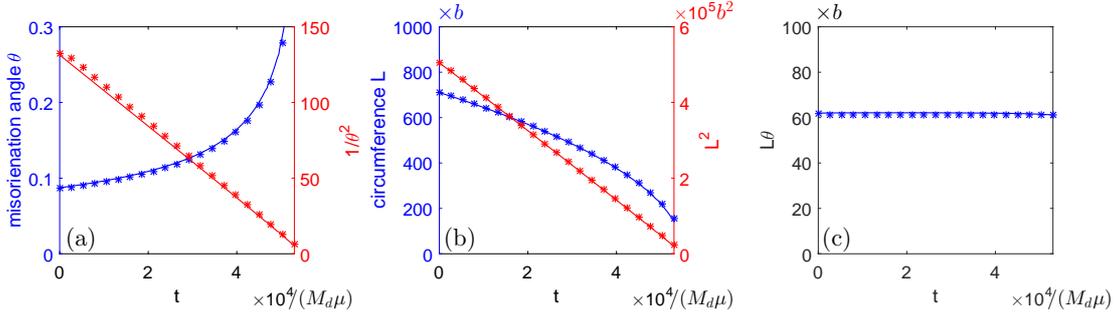}
     \caption{Grain boundary motion under dislocation conservation. The initial grain boundary is an ellipse. (a) The misorientation angle $\theta$ as a function of evolution time. (b) The circumference of the grain boundary $L$  as a function of evolution time. (c) $L\theta$ keeps  constant during the evolution. The curves show the results of the continuum model and the dots show the results using the discrete dislocation model.}
    \label{ellipse_R_rotation}
\end{figure}

Next, we examine the evolution of the misorientation angle $\theta$, which is calculated by using  Eq.~\eqref{eqn:theta_noncircular} during the evolution of the grain boundary.
Fig.~\ref{ellipse_R_rotation}a shows the misorientation angle $\theta$ as a function of the evolution time $t$.
 It can be seen that as in the simulation of a circular grain boundary, $\theta$ increases during the evolution and $1/\theta^2$ is a linear function of $t$. Evolution of the circumference of the elliptic grain boundary, $L$, is shown in Fig.~\ref{ellipse_R_rotation}b. The simulation results show that $L$ is decreasing and $L^2$ is a linear function of $t$, as the behavior of the radius of the circular grain boundary $R$ in the simulations in Sec.~\ref{sec:circle_conservation}. That is
 \begin{flalign}
 L(t)^2=L(0)^2-Bt,
 \end{flalign}
 where $B$ is some constant.
 These results are in excellent agreement with the discrete dislocation dynamics simulations.
  In this case, the quantity $L\theta$ does not change during the evolution, as the quantity $R\theta$ in the simulations of a circular grain in Sec.~\ref{sec:circle_conservation}.
   This also agrees excellently with the discrete dislocation dynamics simulations as shown in Fig.~\ref{ellipse_R_rotation}c and the theoretic prediction in Eq.~\eqref{eqn:theta_noncircular1}.

\subsubsection{With dislocation reaction}\label{sec:ellipse_reaction}

\begin{figure}[htbp]
\centering
    \includegraphics[width=0.6\linewidth]{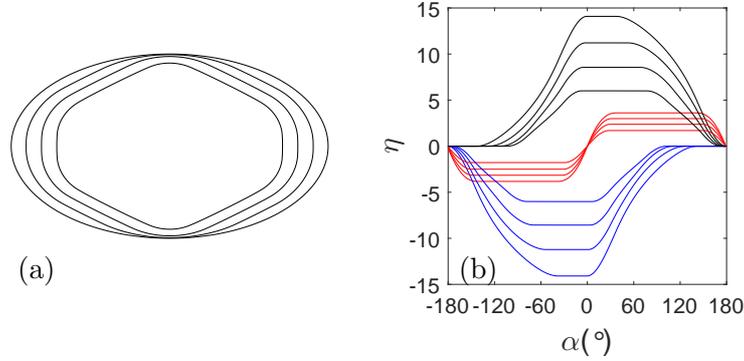}
    \caption{Grain boundary motion with dislocation reaction: $M_{\rm t}b/ M_{\rm d}=0.286\times 10^{-4}$. The initial grain boundary is an ellipse. (a) Grain boundary motion using our continuum model. The grain boundary is plotted at uniform time intervals starting with the outer most one.
     (b) Evolution of the dislocation density potential functions $\eta^{(1)}$, $\eta^{(2)}$, and ${\eta^{(3)}}$ of the three arrays of dislocations with  Burgers vectors $\mathbf{b}^{(1)}$ (red), $\mathbf{b}^{(2)}$ (black), and $\mathbf{b}^{(3)}$ (blue), respectively.    }
    \label{ellipse_motion1}
\end{figure}

In this subsection, we perform simulations from the initially elliptical grain boundary for the case in which dislocation reaction is not negligible during the motion of the grain boundary, i.e., $M_{\rm t}\neq 0$.  The simulation results of the circular grain boundary with $M_{\rm t}b/ M_{\rm d}=0.286\times 10^{-4}$ are shown in Fig.~\ref{ellipse_motion1}a. The initial circular grain boundary gradually changes to hexagonal shape as it shrinks, as the simulations from a circular grain boundary with dislocation reaction presented in Sec.~\ref{sec:circle_reaction}.
Evolution of the dislocation density potential functions $\eta^{(j)}$, $j=1,2,3$, is shown in Fig.~\ref{ellipse_motion1}b. Again as the simulations from a circular grain boundary with dislocation reaction presented in Sec.~\ref{sec:circle_reaction},
the amplitude of each $\eta^{(j)}$ is decreasing,  meaning that the dislocations react and the number of dislocations of each Burgers vector is reduced.

\begin{figure}[htbp]
\centering
     \includegraphics[width=0.9\linewidth]{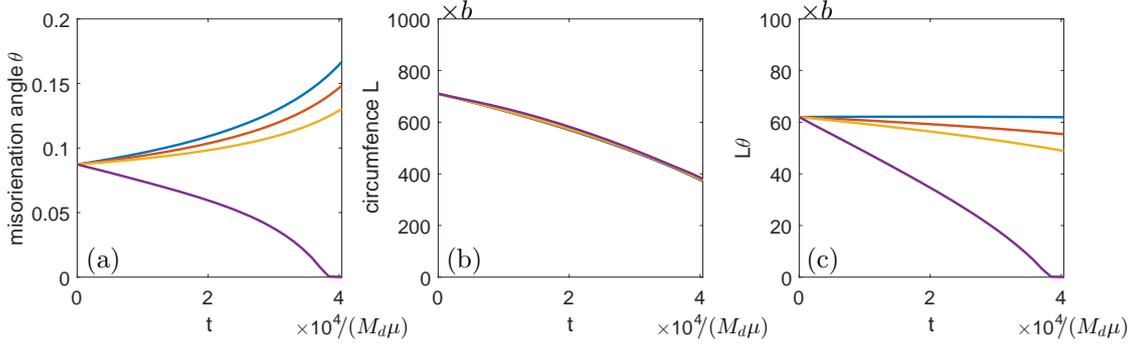}
    \caption{Grain boundary motion with dislocation reaction for different reaction mobility $M_{\rm t}$: $M_{\rm t}b/ M_{\rm d}=0, 2.86\times 10^{-5}, 5.762\times 10^{-5}$, and $2.86\times 10^{-4}$ (from the top to the bottom). The shape of the initial grain boundary is an ellipse.
     (a) Evolution of the misorientation angle $\theta$. (b) Evolution of the circumference  of the grain boundary $L$. (c) Evolution of $L\theta$. }
    \label{ellipse_angle}
\end{figure}

Evolution of the misorietation angle $\theta$ from this initial elliptic grain boundary with different values of the reaction mobility $M_{\rm t}$ is shown in Fig.~\ref{ellipse_angle}a, compared with the evolution of $\theta$ with $M_{\rm t}=0$ discussed previously. As discussed in the simulations of circular grain boundaries,
the coupling effect  that depends on the dislocation mobility $M_{\rm d}$
increases the misorientation angle $\theta$,  while the sliding effect
 that depends mainly on the mobility $M_{\rm t}$ due to dislocation reaction
decreases $\theta$.
When the dislocation reaction mobility $M_{\rm t}$ increases, meaning the sliding effect due to dislocation reaction is becoming stronger, the increase rate of $\theta$ is decreasing during the motion of the grain boundary, and when the sliding effect is strong enough, the misorientation angle $\theta$ is decreasing.

Evolutions of the circumference of the grain boundary $L$ and the product $L\theta$ are shown in Fig.~\ref{ellipse_angle}b and Fig.~\ref{ellipse_angle}c. Again as in the simulations of a circular grain boundary, the product $L\theta$ keeps constant in the case of $M_{\rm t}=0$ due to the pure coupling effect;
whereas with dislocation reaction, the sliding effect is no longer negligible, and $L\theta$ is decreasing during the evolution of the grain boundary due to the fact that the sliding effect decreases the misorientation angle, as shown in Fig.~\ref{ellipse_angle}c.
The evolution of the circumference of the grain boundary $L$ does not change much for these values of $M_{\rm t}$ compared with the case of $M_{\rm t}=0$, see Fig.~\ref{ellipse_angle}b.

\subsection{Pointwise misorientation angle $\theta$}

In the simulations presented above, we calculate the misorientation angle $\theta$ of the grain boundary based on the formula in Eq.~\eqref{eqn:theta_noncircular}, which is an average of the pointwise misorientation angle  along the grain boundary given by Eq.~\eqref{theta_calculation1} based on the Frank's formula.
 In this subsection, we examine
   the pointwise misorientation angle formula in Eq.~\eqref{theta_calculation1} using the simulation results of our continuum model.

\begin{figure}[htbp]
\centering
     \includegraphics[width=0.6\linewidth]{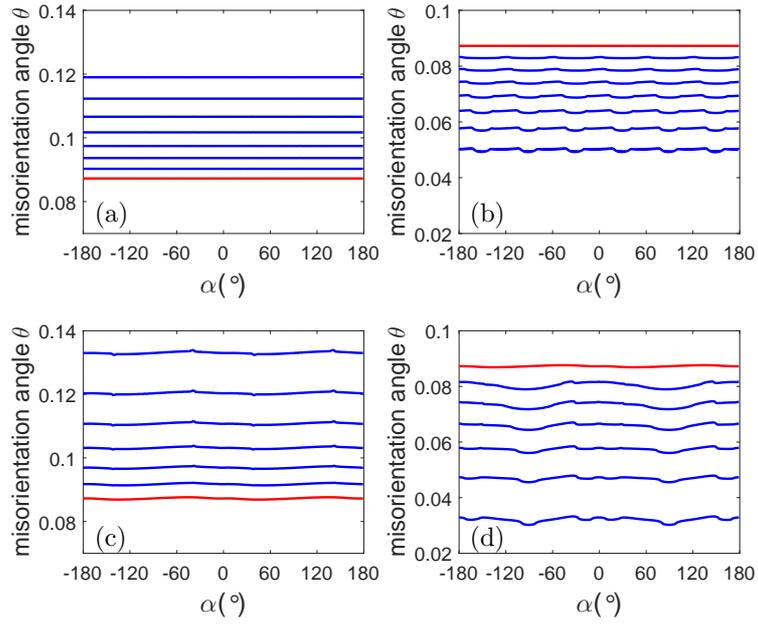}
    \caption{Pointwise misorientation angle $\theta$ during the motion of the grain boundary, plotted at uniform time interval. The red line shows the values of $\theta$ on the initial grain boundary. (a) The shape of the initial grain boundary is a circle, under dislocation conservation. (b) The shape of the initial grain boundary is a circle, with dislocation reaction. (c) The shape of the initial grain boundary is an ellipse,  under dislocation conservation. (d) The shape of the initial grain boundary is an ellipse, with dislocation reaction. In (b) and (d), $M_{\rm t}b/M_{\rm d}=0.286\times10^{-4}$. }
    \label{misorientation_angle}
\end{figure}

Figs.~\ref{misorientation_angle}a and \ref{misorientation_angle}c show the pointwise misorientation angle $\theta$ for the simulation of the motion of the  grain boundary under dislocation conservation for the circular and elliptic grain boundaries presented in Secs.~\ref{sec:circle_conservation} and \ref{sec:ellipse_conservation}, respectively. It can be seen that $\theta$  increases as the grain boundary evolves due to the pure coupling motion.
 In this case, $\theta$ is constant throughout the grain boundary  during its motion, perfectly for the circular grain boundary and approximately for the elliptic one.
 Figs.~\ref{misorientation_angle}b and \ref{misorientation_angle}d show the calculated pointwise value of  $\theta$ for the simulation of the motion of the grain boundary with dislocation reaction  for the intitially circular and elliptic grain boundaries presented in Secs.~\ref{sec:circle_reaction} and \ref{sec:ellipse_reaction}, respectively,
 where the mobility $M_{\rm t}$ associated with dislocation reaction is relatively large and  $\theta$  decreases as the grain boundary evolves due to the strong sliding effect.
  In this case, $\theta$ is also a constant approximately throughout the grain boundary  during its motion. Note that the grain boundary shape is changed once it evolves.
 The small deviations from a perfect constant of the pointwise misorientation angle $\theta$ on the grain boundary in  Figs.~\ref{misorientation_angle}b and d are due to the fact that the long-range elastic interaction between the constituent dislocations of the grain boundary that maintains the Frank's formula is finite, thus is not able to make the Frank's formula hold instantly after the dislocation reactions during the evolution.

These numerical results show that the pointwise misorientation angle formula in Eq.~\eqref{theta_calculation1} indeed holds throughout the grain boundary during its evolution. Since Eq.~\eqref{theta_calculation1} is derived from the Frank's formula, these simulation results also provide
 a justification that the Frank's formula holds pointwise during the motion of the grain boundary, which is the basis for  the formulas in Eqs.~\eqref{theta_calculation1} and \eqref{eqn:theta_noncircular} for the misorientation angle (pointwise and averaged, respectively), Eq.~\eqref{eqn:dtheta_dt} for the grain rotation, and Eq.~\eqref{eqn:coupling0} for the tangential velocity of the grain boundary.

\section{Conclusions and discussion}\label{sec:conclusion}

In this paper, we present a continuum model for the dynamics of low angle grain boundaries in two dimensions based on the motion of constituent dislocations of the grain boundaries. The continuum model consists of an equation for the motion of the grain boundary that describes the motion of the constituent dislocations in the grain boundary normal direction, and equations for the dislocation structure evolution on the grain boundary (Eqs.~\eqref{eqn:vn} and \eqref{eqn:vp}, or Eqs.~\eqref{eqn:vr} and \eqref{eqn:vpr} in Sec.~\ref{sec:model}).
The long-range elastic interaction between dislocations is included in the continuum model, which maintains the dislocation structures on the grain boundaries and ensures that they are consistent with the condition of Frank's formula for grain boundaries, i.e., the condition of cancellation of the far-field elastic fields.
These evolutions of the grain boundary and its dislocation structure are able to describe both motion and tangential translation of grain boundaries and grain rotation due to both coupling and sliding.
Since the continuum model is based upon dislocation structure, it naturally accounts for the grain boundary shape change during the motion and rotation of the grain boundary by motion and reaction of the constituent dislocations without explicit mass transfer. See Eqs.~\eqref{theta_calculation1} and \eqref{eqn:theta_noncircular} for the pointwise and averaged misorientation angle formulas, Eq.~\eqref{eqn:dtheta_dt} for the grain rotation formula, and Eq.~\eqref{eqn:coupling0} for the tangential velocity formula, derived using this continuum grain boundary dynamics model.

Our continuum grain boundary dynamics model is based upon the continuum framework for grain boundaries in \citet{Zhu2014175} derived from the discrete dislocation dynamics model, which is the basis for the development of grain boundary models and understanding of  simulation and experimental results for grain boundaries \citep{ReadShockley1950,Li1953223,Cahn20021,Cahn20044887,Cahn20064953,
Cahn20063965,Molodov20071843,Molodov20095396,Trautt20122407,Wu2012407,Voorhees2016264,Voorhees2017,Lim2009,Quek2011,Lim2012}.

 Compared with the theory of   Cahn-Taylor \citep{Cahn20044887}  in which the coupling effect is an assumption, our model is based on the motion of the grain boundary dislocations and the coupling effect is a result.
 Our continuum model generalizes the theory of   Cahn-Taylor
 by  (1) incorporating detailed formulas of the driving forces for the normal and tangential grain boundary velocities that depend on the constituent dislocations, their Burgers vectors, and the grain boundary shape,
  and (2) incorporating the  shape change of the grain boundaries. Our model is different from their earlier generalized model   based on mass transfer via surface diffusion \citep{Taylor2007493}.
  Compared with the existing continuum models for the motion of grain boundaries and grain rotation based on  evolution of the misorientation angle \citep{Li1962,Shewmon1966,Kobayashi2000,Upmanyu2006,Selim2016}, our dislocation-structure based continuum model is able to further describe the grain rotation due to the coupling of normal and tangential motions of the grain boundaries, which is missing in these existing models.

 Using the derived continuum grain boundary dynamic model, simulations are performed for the dynamics of circular and non-circular two dimensional grain boundaries, and the results are validated  by discrete dislocation dynamics simulations. Simulations of both continuum and discrete models show that Frank's formula holds approximately during the evolution of the grain boundary, and is maintained by the long-range elastic interaction of the constituent dislocations. For a finite grain embedded in another one with a low misorientation angle $\theta$, the coupling motion increases $\theta$ while the sliding motion reduces $\theta$.
Simulations also show that when the grain boundary motion is pure coupling, its shape is preserved (which agrees with prediction of the model of \citep{Taylor2007493} based on mass transfer via surface diffusion),  and the product $L\theta$ keeps constant, where $L$ is the circumference of the grain boundary. The product
 $L\theta$ is decreasing when the sliding effect is not negligible during the motion of the grain boundary. In general, a grain boundary is not able to maintain a circular shape during its evolution except for some special cases.

The continuum model for the dynamics of low angle grain boundary presented in this paper can be generalized to multiple low angle grain boundaries with junctions. Generalizations  can also be made to grain boundaries in three dimensions. These will be explored in the future work. Finally, we remark that
continuum model for the dynamics of high angle grain boundaries has also been developed \citep{ZhangLuchan2} based on a disconnection model and atomistic simulations \citep{Thomas2017}.

\appendix
\section{Parametrization of the grain boundary using polar angle $\alpha$}\label{sec:geometryformulas}

 We parametrize the grain boundary by the polar angle $\alpha$ as shown in Fig.~\ref{fig:geometry}. A point on the grain boundary $\Gamma$ has the coordinate
$\mathbf R(\alpha)= (x(\alpha),y(\alpha))=(R(\alpha)\cos\alpha,R(\alpha)\sin\alpha)$.
   The grain boundary tangent direction $\mathbf T$, the grain boundary normal direction $\mathbf n$,  and the curvature of the grain boundary $\kappa$ at a point $(x(\alpha),y(\alpha))$ on the grain boundary can be calculated as
\begin{flalign}
\mathbf T&={\textstyle \frac{1}{\sqrt{{x'(\alpha)}^2+{y'(\alpha)}^2}}\left(x'(\alpha),y'(\alpha)\right)}, \label{eqn:tangent}
\end{flalign}
\begin{flalign}
\mathbf n&={\textstyle \frac{1}{\sqrt{{x'(\alpha)}^2+{y'(\alpha)}^2}}\left(-y'(\alpha),x'(\alpha)\right)}, \label{eqn:normal}
\end{flalign}
\begin{flalign}
\kappa&= \frac{R(\alpha)^2+2R'(\alpha)^2-R''(\alpha)R(\alpha)}{\left[R(\alpha)^2+R'(\alpha)^2\right]^{3/2}}. \label{eqn:curvature}
\end{flalign}
The normal direction of the grain boundary $\mathbf n$ in Eq.~\eqref{eqn:normal} is defined such that $\mathbf T\times \mathbf n=\mathbf k$, where $\mathbf k$ is the unit vector in the $+z$ direction. We have the relation
\begin{flalign}
 \frac{d\mathbf T}{d s}=\kappa, \ \
 \frac{d\mathbf n}{d s}=-\kappa \mathbf T,
 \end{flalign}
where $s$ is the arclength parameter of the grain boundary $\Gamma$.
 We also have the relation
 \begin{flalign}
 \frac{d}{d\alpha}=\frac{ds}{d\alpha}\frac{d}{ds},
 \end{flalign}
  and
 \begin{flalign}
 \frac{ds}{d\alpha}=\sqrt{x'(\alpha)^2+y'(\alpha)^2}=\sqrt{R(\alpha)^2+R'(\alpha)^2}. \label{eqn:arclength}
  \end{flalign}

When the grain boundary is evolved with a velocity in the radial direction $\partial R/\partial t$, the evolution of the normal direction of the grain boundary is
\begin{flalign}\label{eqn:dndt}
\frac{\partial \mathbf n}{\partial t}=\frac{\partial }{\partial t}\tan^{-1}\left(\frac{x'}{-y'}\right)\mathbf T
=\frac{R\frac{\partial R'}{\partial t}-R'\frac{\partial R}{\partial t}}{R^2+R'^2}\mathbf T.
\end{flalign}

\section{Variational derivation of the continuum model for grain boundary motion}\label{sec:variation}

In the classical theories for the motion of grain boundaries, the driving force is obtained by variation of the grain boundary energy whose density does not evolve as the grain boundary migrates \citep{Herring1951,Mullins1956,Sutton1995}. Although in some models, further variations have been taken to reduce the grain boundary energy density (grain boundary sliding) \citep{Li1962,Kobayashi2000,Upmanyu2006,Selim2016}, these models are not able to include the grain boundary coupling motion in which the grain boundary energy density may increase \citep{Cahn20021,Cahn20044887}.

In the continuum framework for grain boundaries and dislocation arrays proposed in \citet{Zhu2014175}, the driving force for grain boundary or dislocation array motion was obtained by taking variation of the total energy   with respect to the change of the grain boundary profile under the conservation of the constituent dislocations. This enables the incorporation of both the coupling and sliding motions of the grain boundary when it evolves. This continuum framework is general: it applies to dislocation structures on grain boundaries in three dimensions and includes both the long-range and short-range interactions of the constituent dislocations.

In this section of Appendix, we present an example of such variation method for the derivation of the continuum grain boundary motion model in two dimensions,  in terms of the contribution of the  grain boundary energy due to the local interaction of dislocations which is the commonly used grain boundary energy in the literature. This approach is a generalization of that by \citet{Cahn20021} and  \citet{Cahn20044887} for two dimensional circular grain boundaries, and is different from the model by Talor-Cahn based on explicit mass transfer along the grain boundaries \citep{Taylor2007493}.

Consider a grain boundary $\Gamma$ as shown in Fig.~\ref{fig:geometry}. For the purpose of demonstration of the method, we consider the simple case where all the dislocations have the same Burgers vector $\mathbf b$ which is also in the $xy$ plane. The local grain boundary energy in Eq.~\eqref{eqn:gb_density} in this case is,
\begin{flalign}
E_{\rm local}&=\int_\Gamma \gamma ds,\vspace{1ex}\label{eqn:gb_energyA}\\
\gamma&= \frac{\mu b^2}{4\pi(1-\nu)}|\rho|
\log\! \frac{1}{r_g|\rho|}, \label{eqn:gb_densityA}
\end{flalign}
using the dislocation density per unit length $\rho$.

The variation with respect to the change of $\Gamma$ is taken under the conservation of the constituent dislocations. Assuming that there is a small change in the location of the grain boundary $\Gamma$ in its normal direction $\mathbf n$, denoted by $\delta r \mathbf n$,  the change of energy is
\begin{flalign}\label{eqn:delta_E}
\delta E_{\rm local}=\delta\left(\int_\Gamma \gamma ds\right)
=\int_\Gamma \delta\left(\gamma\right) ds+\int_\Gamma \gamma \! \delta(ds).
\end{flalign}
In the classical theory of motion by curvature \citep{Herring1951,Mullins1956,Sutton1995}, it has been shown that the change of a small arclength $ds$ is
\begin{flalign}\label{eqn:delta_ds}
\delta(ds)=-\kappa \delta r ds,
\end{flalign}
where $\kappa$ is the curvature of the grain boundary.

 For the small arclength $ds$, the total number of dislocations along $ds$ is $|\rho| ds$. (Note: Using the dislocation density potential function $\eta$, it is $d\eta=|\rho| ds$, which is very convenient in the derivation in three dimensions \citep{Zhu2014175}.) Conservation of the constituent dislocations gives
\begin{flalign}
\delta(|\rho| ds)=\delta|\rho|\! ds+|\rho| \delta(ds)=0.
\end{flalign}
Further using Eq.~\eqref{eqn:delta_ds}, we have $\delta|\rho|\! ds+|\rho| (-\kappa \delta r ds)=0$, or
\begin{flalign}
\delta|\rho|=\kappa|\rho|\delta r.
\end{flalign}
Using the energy density formula in Eq.~\eqref{eqn:gb_densityA}, it can be calculated that
\begin{flalign}\label{eqn:delta_gamma}
\delta\left(\gamma\right)=\frac{ d\gamma}{d |\rho|}\delta|\rho|
=\frac{ d\gamma}{d |\rho|}\kappa|\rho|\delta r.
\end{flalign}
Substituting Eqs.~\eqref{eqn:delta_ds} and \eqref{eqn:delta_gamma} into Eq.~\eqref{eqn:delta_E}, we have
\begin{flalign}
\delta E_{\rm local}
=-\int_\Gamma \kappa \left(\gamma -|\rho|\frac{ d\gamma}{d |\rho|} \right)\delta r ds.
\end{flalign}
Therefore, we have the variation of the energy with respect to the change of the grain boundary profile:
\begin{flalign}\label{eqn:variation_delta_r}
\frac{\delta E_{\rm local}}{\delta r}
=- \kappa \left(\gamma -|\rho|\frac{ d\gamma}{d |\rho|} \right)=-  \frac{\mu b^2}{4\pi(1-\nu)}\kappa |\rho|.
\end{flalign}
This is a special case of the variation obtained in \citet{Zhu2014175}, see Eq.~\eqref{eqn:var_n} with \eqref{eqn:fs} in Sec.~\ref{sec:settings}.

Therefore, the normal velocity of the grain boundary due to the driving force of the variation of the local grain boundary energy is
\begin{flalign}
v_n=-\frac{M_{\rm d}}{|\rho|}\frac{\delta E_{\rm local}}{\delta r}
=  M_{\rm d}\frac{\mu b^2}{4\pi(1-\nu)}\kappa,
\end{flalign}
where $M_{\rm d}$ is the mobility of the constituent dislocations of the grain boundary.
Here a factor $1/|\rho|$ has been included with the dislocation mobility $M_{\rm d}$  so that the velocity of the grain boundary is  consistent with the dynamics of its constituent dislocations \citep{Cahn20044887}. This can be seen more explicitly from the full variation formula in Eqs.~\eqref{eqn:var_n} and \eqref{eqn:fs} with all types of driving forces.

Next we consider the variation of the local grain boundary  energy in Eqs.~\eqref{eqn:gb_energyA} and \eqref{eqn:gb_densityA} with respect to the change of the dislocation density per unit polar angle $\varrho$ on the fixed grain boundary. It is easy to calculate that
\begin{flalign}
\frac{\delta E_{\rm local}}{\delta \varrho}
=\frac{ d\gamma}{d \varrho}.
\end{flalign}
Therefore, the evolution of the dislocation density $\varrho$ on the  grain boundary due to this variation is
\begin{flalign}
\varrho_t=-M_{\rm t}\frac{\delta E_{\rm local}}{\delta \varrho}
=-M_{\rm t}\frac{ d\gamma}{d \varrho},
\end{flalign}
where $M_{\rm t}$ is the mobility.

\section{Proof of equivalence of Frank's formula and cancellation of the  long-range elastic fields}\label{sec:equivalenceproof}

A proof of the equivalence of the Frank's formula and cancellation of the  long-range elastic fields  generated by the constituent dislocations for a curved grain boundary can be found in \citet{Zhu2014175}. Here we present an alternative calculation in two dimensions to show their equivalence. In fact,
the long-range stress field  generated by the constituent dislocations of the grain boundary $\Gamma$, in the infinite two dimensional space, can be written as
\begin{flalign}\label{generalstress}
\pmb\sigma_{\rm i}(x,y)=&\sum_{j=1}^J\int_{\Gamma}\left(\mathbf G_1(x,y;x_1,y_1)b^{(j)}_1\rho^{(j)}(x_1,y_1)+\mathbf G_2(x,y;x_1,y_1)b^{(j)}_2\rho^{(j)}(x_1,y_1)\right)ds,
\end{flalign}
where the point $(x_1,y_1)$ varies along the grain boundary in the integrals, $ds$ is the line element of the integrals, and
\begin{flalign}
\mathbf G_1(x,y;x_1,y_1)&={\textstyle \frac{\mu}{2\pi(1-\nu)}}
\left(
\begin{array}{cc}\label{G1}
-\frac{(y-y_1)[3(x-x_1)^2+(y-y_1)^2]}{[(x-x_1)^2+(y-y_1)^2]^2}&\frac{(x-x_1)[(x-x_1)^2-(y-y_1)^2]}{[(x-x_1)^2+(y-y_1)^2]^2}\vspace{1ex}\\
\frac{(x-x_1)[(x-x_1)^2-(y-y_1)^2]}{[(x-x_1)^2+(y-y_1)^2]^2}&\frac{(y-y_1)[(x-x_1)^2-(y-y_1)^2]}{[(x-x_1)^2+(y-y_1)^2]^2}
\end{array}
\right),\\
\mathbf G_2(x,y;x_1,y_1)&={\textstyle \frac{\mu}{2\pi(1-\nu)}}
\left(
\begin{array}{cc}
\frac{(x-x_1)[(x-x_1)^2-(y-y_1)^2]}{[(x-x_1)^2+(y-y_1)^2]^2}&\frac{(y-y_1)[(x-x_1)^2-(y-y_1)^2]}{[(x-x_1)^2+(y-y_1)^2]^2}\vspace{1ex}\\
\frac{(y-y_1)[(x-x_1)^2-(y-y_1)^2]}{[(x-x_1)^2+(y-y_1)^2]^2}&\frac{(x-x_1)[(x-x_1)^2+3(y-y_1)^2]}{[(x-x_1)^2+(y-y_1)^2]^2}
\end{array}
\right).\label{G2}
\end{flalign}

Using divergence theorem and the fact that $\frac{\partial}{\partial x_1}\mathbf G_1+ \frac{\partial}{\partial y_1}\mathbf G_2=\mathbf 0$, we have
\begin{flalign}\label{Gdivergence}
\int_{\Gamma}\left(\mathbf G_1\theta n_1+\mathbf G_2\theta n_2\right)ds
=\theta \int_\Omega \left(\frac{\partial \mathbf G_1}{\partial x_1}+ \frac{\partial \mathbf G_2}{\partial y_1}\right)dx_1dy_1=\mathbf 0,
\end{flalign}
where  $\mathbf n=(n_1,n_2)$ is recalled to be the unit normal vector of the grain boundary and $\Omega$ is the inner grain.
Using Eqs.~\eqref{generalstress} and \eqref{Gdivergence}, the stress field due to the long-range dislocation interaction can be written as
\begin{flalign}\label{generalstress2}
\pmb\sigma_{\rm i}(x,y)=&\sum_{j=1}^J\int_{\Gamma}\left[\mathbf G_1(x,y;x_1,y_1)\left(b^{(j)}_1\rho^{(j)}(x_1,y_1)-\theta n_1(x_1,y_1)\right)\right.
\vspace{1ex}\\
&+\left.\mathbf G_2(x,y;x_1,y_1)\left(b^{(j)}_2\rho^{(j)}(x_1,y_1)-\theta n_2(x_1,y_1)\right)\right]ds. \nonumber
\end{flalign}
Therefore, the long-range stress field $\pmb\sigma_{\rm i}$, and accordingly the long-range elastic interaction energy of the constituent dislocations, vanish when the Frank's formula in Eq.~\eqref{eqn:frank1} holds.

\section{Derivation of the grain rotation formula in Eq.~\eqref{eqn:dtheta_dt}}\label{eqn:grainrotationderivation}

The grain rotation formula in Eq.~\eqref{eqn:dtheta_dt} is derived as follows. Taking time derivative in Eq.~\eqref{eqn:frank2}, we have
\begin{flalign}\label{eqn:dFrankdt}
\frac{d\theta}{dt}\mathbf n+\theta\frac{\partial \mathbf n}{\partial t}
+\sum_{j=1}^J \varrho^{(j)}\mathbf b^{(j)}\frac{\partial }{\partial t}\left(\frac{1}{\sqrt{R^2+R'^2}}\right)
+\sum_{j=1}^J \frac{\mathbf b^{(j)}}{\sqrt{R^2+R'^2}}\frac{\partial  \varrho^{(j)}}{\partial t}=\mathbf 0.
\end{flalign}
Taking dot product with $\mathbf n$ and $\mathbf T$, respectively, in Eq.~\eqref{eqn:dFrankdt} and using Eq.~\eqref{eqn:dndt} and the Frank's formula in Eq.~\eqref{eqn:frank2}, we have
\begin{flalign}
\frac{d\theta}{dt}=-\theta\frac{R\frac{\partial R}{\partial t}+R'\frac{\partial R'}{\partial t}}{R^2+R'^2}
-\sum_{j=1}^J \frac{\mathbf b^{(j)}\cdot\mathbf n}{\sqrt{R^2+R'^2}}\frac{\partial \varrho^{(j)}}{\partial t}\\
\theta\frac{R\frac{\partial R'}{\partial t}-R'\frac{\partial R}{\partial t}}{R^2+R'^2}=
-\sum_{j=1}^J \frac{\mathbf b^{(j)}\cdot\mathbf T}{\sqrt{R^2+R'^2}}\frac{\partial \varrho^{(j)}}{\partial t}.
\end{flalign}
The second equation gives
\begin{flalign}
\frac{\partial R'}{\partial t}=\frac{R'}{R}\frac{\partial R}{\partial t}
-\sum_{j=1}^J \frac{\sqrt{R^2+R'^2}}{\theta R}\left(\mathbf b^{(j)}\cdot\mathbf T\right)\frac{\partial \varrho^{(j)}}{\partial t},
\end{flalign}
and inserting it into the first equation, we have
\begin{flalign}
\frac{d\theta}{dt}
=-\frac{\theta}{R}\frac{\partial R}{\partial t}
+\frac{1}{\sqrt{R^2+R'^2}}\sum_{j=1}^J \left(\frac{R'}{R}\mathbf b^{(j)}\cdot\mathbf T-\mathbf b^{(j)}\cdot\mathbf n\right)\frac{\partial \varrho^{(j)}}{\partial t}.
\end{flalign}
This gives Eq.~\eqref{eqn:dtheta_dt} by using the equation that $\frac{R'}{\sqrt{R^2+R'^2}}\mathbf T-\frac{R}{\sqrt{R^2+R'^2}}\mathbf n=\hat{\mathbf R}$, which
 can be obtained using the formulas of $\mathbf T$ and $\mathbf n$ in Eqs.~\eqref{eqn:tangent} and  \eqref{eqn:normal}.

\section*{Acknowledgments}
 This work was partially supported by the Hong Kong Research Grants Council General Research Fund 606313.

\bibliography{science_v4_rev}

\begin{thebibliography}{46}
\providecommand{\natexlab}[1]{#1}
\providecommand{\url}[1]{\texttt{#1}}
\expandafter\ifx\csname urlstyle\endcsname\relax
  \providecommand{\doi}[1]{doi: #1}\else
  \providecommand{\doi}{doi: \begingroup \urlstyle{rm}\Url}\fi

\bibitem[Basak and Gupta(2014)]{Gupta2014}
A.~Basak and A.~Gupta.
\newblock A two-dimensional study of coupled grain boundary motion using the
  level set method.
\newblock \emph{Modell. Simul. Mater. Sci. Eng.}, 22:\penalty0 055022, 2014.

\bibitem[Bilby(1955)]{Bilby}
B.~A. Bilby.
\newblock In \emph{Bristol conference report on defects in crystalline
  materials}, page 123. Physical Society, London, 1955.

\bibitem[Cahn and Taylor(2004)]{Cahn20044887}
J.~W. Cahn and J.~E. Taylor.
\newblock A unified approach to motion of grain boundaries, relative tangential
  translation along grain boundaries, and grain rotation.
\newblock \emph{Acta Mater.}, 52:\penalty0 4887--4898, 2004.

\bibitem[Cahn et~al.(2006{\natexlab{a}})Cahn, Mishin, and Suzuki]{Cahn20063965}
J.~W. Cahn, Y.~Mishin, and A.~Suzuki.
\newblock Duality of dislocation content of grain boundaries.
\newblock \emph{Philos. Mag.}, 86:\penalty0 3965--3980, 2006{\natexlab{a}}.

\bibitem[Cahn et~al.(2006{\natexlab{b}})Cahn, Mishin, and Suzuki]{Cahn20064953}
J.~W. Cahn, Y.~Mishin, and A.~Suzuki.
\newblock Coupling grain boundary motion to shear deformation.
\newblock \emph{Acta Mater.}, 54:\penalty0 4953--4975, 2006{\natexlab{b}}.

\bibitem[Chen and Yang(1994)]{Chenlq1994}
L.~Q. Chen and W.~Yang.
\newblock Computer simulation of the domain dynamics of a quenched system with
  a large number of nonconserved order parameters: The grain-growth kinetics.
\newblock \emph{Phys. Rev. B}, 50:\penalty0 15752--15756, 1994.

\bibitem[Elsey et~al.(2009)Elsey, Esedoglu, and Smereka]{Selim2009}
M.~Elsey, S.~Esedoglu, and P.~Smereka.
\newblock Diffusion generated motion for grain growth in two and three
  dimensions.
\newblock \emph{J. Comput. Phys.}, 228:\penalty0 8015--8033, 2009.

\bibitem[Esedoglu(2016)]{Selim2016}
S.~Esedoglu.
\newblock Grain size distribution under simultaneous grain boundary migration
  and grain rotation in two dimensions.
\newblock \emph{Comput. Mater. Sci.}, 121:\penalty0 209--216, 2016.

\bibitem[Frank(1950)]{Frank}
F.~C. Frank.
\newblock The resultant content of dislocations in an arbitrary
  intercrystalline boundary.
\newblock In \emph{Symposium on the plastic deformation of crystalline solids},
  pages 150--154. Office of Naval Research, Pittsburgh, 1950.

\bibitem[Gorkaya et~al.(2009)Gorkaya, Molodov, and Gottstein]{Molodov20095396}
T.~Gorkaya, D.~A. Molodov, and G.~Gottstein.
\newblock Stress-driven migration of symmetrical $<100>$ tilt grain boundaries
  in al bicrystals.
\newblock \emph{Acta Mater.}, 57:\penalty0 5396--5405, 2009.

\bibitem[Harris et~al.(1998)Harris, Singh, and King]{Harris19982623}
K.~E. Harris, V.~V. Singh, and A.~H. King.
\newblock Grain rotation in thin films of gold.
\newblock \emph{Acta Mater.}, 46:\penalty0 2623 -- 2633, 1998.

\bibitem[Herring(1951)]{Herring1951}
C.~Herring.
\newblock Surface tension as a motivation for sintering.
\newblock In W.~E. Kingston, editor, \emph{The Physics of Powder Metallurgy},
  pages 143--179. McGraw-Hill, New York, 1951.

\bibitem[Hirth and Lothe(1982)]{HL}
J.~P. Hirth and J.~Lothe.
\newblock \emph{Theory of Dislocations}.
\newblock Wiley, New York, second edition, 1982.

\bibitem[Kazaryan et~al.(2000)Kazaryan, Wang, Dregia, and Patton]{Kazaryan2000}
A.~Kazaryan, Y.~Wang, S.~A. Dregia, and B.~R. Patton.
\newblock Generalized phase-field model for computer simulation of grain growth
  in anisotropic systems.
\newblock \emph{Phys. Rev. B}, 61:\penalty0 14275--14278, 2000.

\bibitem[Kirch et~al.(2006)Kirch, Jannot, Barrales-Mora, Molodov, and
  Gottstein]{Kirch2006}
D.~M. Kirch, E.~Jannot, L.~A. Barrales-Mora, D.~A. Molodov, and G.~Gottstein.
\newblock Inclination dependence of grain boundary energy and its impact on the
  faceting and kinetics of tilt grain boundaries in aluminum.
\newblock \emph{Acta Mater.}, 56:\penalty0 4998--5011, 2006.

\bibitem[Kobayashi et~al.(2000)Kobayashi, Warren, and Carter]{Kobayashi2000}
R.~Kobayashi, J.~A. Warren, and W.~C. Carter.
\newblock A continuum model of grain boundaries.
\newblock \emph{Phys. D}, 140:\penalty0 141--150, 2000.

\bibitem[Lazar et~al.(2010)Lazar, MacPherson, and Srolovitz]{Srolovitz2010}
E.~A. Lazar, R.~D. MacPherson, and D.~J. Srolovitz.
\newblock A more accurate two-dimensional grain growth algorithm.
\newblock \emph{Acta Mater.}, 58:\penalty0 364--372, 2010.

\bibitem[Li et~al.(1953)Li, Edwards, Washburn, and Parker]{Li1953223}
C.~H. Li, E.~H. Edwards, J.~Washburn, and E.~R. Parker.
\newblock Stress-induced movement of crystal boundaries.
\newblock \emph{Acta Metall.}, 1:\penalty0 223--229, 1953.

\bibitem[Li(1962)]{Li1962}
J.~C.~M. Li.
\newblock Possibility of subgrain rotation during recrystallization.
\newblock \emph{J. Appl. Phys.}, 33:\penalty0 2958--2965, 1962.

\bibitem[Lim et~al.(2009)Lim, Srolovitz, and Haataja]{Lim2009}
A.~T. Lim, D.~J. Srolovitz, and M.~Haataja.
\newblock Low-angle grain boundary migration in the presence of extrinsic
  dislocations.
\newblock \emph{Acta Mater.}, 57:\penalty0 5013--5022, 2009.

\bibitem[Lim et~al.(2012)Lim, Haataja, Cai, and Srolovitz]{Lim2012}
A.~T. Lim, M.~Haataja, W.~Cai, and D.~J. Srolovitz.
\newblock Stress-driven migration of simple low-angle mixed grain boundaries.
\newblock \emph{Acta Mater.}, 60:\penalty0 1395--1407, 2012.

\bibitem[McReynolds et~al.(2016)McReynolds, Wu, and Voorhees]{Voorhees2016264}
K.~McReynolds, K.~A Wu, and P.~Voorhees.
\newblock Grain growth and grain translation in crystals.
\newblock \emph{Acta Mater.}, 120:\penalty0 264--272, 2016.

\bibitem[Molodov et~al.(2007)Molodov, Ivanov, and Gottstein]{Molodov20071843}
D.~A. Molodov, V.~A. Ivanov, and G.~Gottstein.
\newblock Low angle tilt boundary migration coupled to shear deformation.
\newblock \emph{Acta Mater.}, 55:\penalty0 1843--1848, 2007.

\bibitem[Mullins(1956)]{Mullins1956}
W.~Mullins.
\newblock Two-dimensional motion of idealized grain boundaries.
\newblock \emph{J. Appl. Phys.}, 27:\penalty0 900--904, 1956.

\bibitem[Quek et~al.(2011)Quek, Xiang, and Srolovitz]{Quek2011}
S.~S. Quek, Y.~Xiang, and D.~J. Srolovitz.
\newblock Loss of interface coherency around a misfitting spherical inclusion.
\newblock \emph{Acta Mater.}, 59:\penalty0 5398--5410, 2011.

\bibitem[Rath et~al.(2007)Rath, Winning, and Li]{Rath2007}
B.~B. Rath, M.~Winning, and J.~C.~M. Li.
\newblock Coupling between grain growth and grain rotation.
\newblock \emph{Appl. Phys. Lett.}, 90:\penalty0 161915, 2007.

\bibitem[Read and Shockley(1950)]{ReadShockley1950}
W.~T. Read and W.~Shockley.
\newblock Dislocation models of crystal grain boundaries.
\newblock \emph{Phys. Rev.}, 75:\penalty0 275--289, 1950.

\bibitem[Shewmon(1966)]{Shewmon1966}
P.~G. Shewmon.
\newblock In H.~Margolin, editor, \emph{Recrystallization, grain growth and
  textures}, pages 165--199. American Society of Metals, Metals Park, 1966.

\bibitem[Sidi and Israeli(1988)]{SI1988}
A.~Sidi and M.~Israeli.
\newblock Quadrature methodsforperiodic singular and weakly singular fredholm
  integral equations.
\newblock \emph{J. Sci. Comput.}, 3:\penalty0 201--231, 1988.

\bibitem[Srinivasan and Cahn(2002)]{Cahn20021}
S.~G. Srinivasan and J.~W. Cahn.
\newblock Challenging some free-energy reduction criteria for grain growth.
\newblock In S.~Ankem, C.~S. Pande, I.~Ovid'ko, and S.~Ranganathan, editors,
  \emph{Science and Technology of Interfaces}, pages 3--14. TMS, Seattle, 2002.

\bibitem[Sutton and Balluffi(1995)]{Sutton1995}
A.P. Sutton and R.W. Balluffi.
\newblock \emph{Interfaces in Crystalline Materials}.
\newblock Clarendon Press, Oxford, 1995.

\bibitem[Taylor and Cahn(2007)]{Taylor2007493}
J.~E. Taylor and J.~W. Cahn.
\newblock Shape accommodation of a rotating embedded crystal via a new
  variational formulation.
\newblock \emph{Interfaces and Free Boundaries}, 9:\penalty0 493--512, 2007.

\bibitem[Thomas et~al.(2017)Thomas, Chen, Han, Purohit, and
  Srolovitz]{Thomas2017}
S.~L. Thomas, K.~T. Chen, J.~Han, P.~K. Purohit, and D.~J. Srolovitz.
\newblock Reconciling grain growth and shear-coupled grain boundary migration.
\newblock \emph{Nature Commun.}, 8:\penalty0 1764, 2017.

\bibitem[Trautt and Mishin(2012)]{Trautt20122407}
Z.~T. Trautt and Y.~Mishin.
\newblock Grain boundary migration and grain rotation studied by molecular
  dynamics.
\newblock \emph{Acta Mater.}, 60:\penalty0 2407--2424, 2012.

\bibitem[Upmanyu et~al.(1998)Upmanyu, Smith, and Srolovitz]{Upmanyu1998}
M.~Upmanyu, R.~W. Smith, and D.~J. Srolovitz.
\newblock Atomistic simulation of curvature driven grain boundary migration.
\newblock \emph{Interface Sci.}, 6:\penalty0 41--58, 1998.

\bibitem[Upmanyu et~al.(2002)Upmanyu, Hassold, Kazaryan, Holm, Wang, Patton,
  and Srolovitz]{Upmanyu2002}
M.~Upmanyu, G.~N. Hassold, A.~Kazaryan, E.~A. Holm, Y.~Wang, B.~Patton, and
  D.~J. Srolovitz.
\newblock Boundary mobility and energy anisotropy effects on microstructural
  evolution during grain growth.
\newblock \emph{Interface Sci.}, 10:\penalty0 201--216, 2002.

\bibitem[Upmanyu et~al.(2006)Upmanyu, Srolovitz, Lobkovsky, Warren, and
  Carter]{Upmanyu2006}
M.~Upmanyu, D.~J. Srolovitz, A.~E. Lobkovsky, J.~A. Warren, and W.~C. Carter.
\newblock Simultaneous grain boundary migration and grain rotation.
\newblock \emph{Acta Mater.}, 54:\penalty0 1707--1719, 2006.

\bibitem[Wu and Voorhees(2012)]{Wu2012407}
K.~A. Wu and P.~W. Voorhees.
\newblock Phase field crystal simulations of nanocrystalline grain growth in
  two dimensions.
\newblock \emph{Acta Mater.}, 60:\penalty0 407--419, 2012.

\bibitem[Xiang and Yan(2017)]{Xiang-Yan2017}
Y.~Xiang and X.~D. Yan.
\newblock Stability of dislocation networks of low angle grain boundaries using
  a continuum energy formulation.
\newblock \emph{Dis. Cont. Dyn. Sys. B}, doi:10.3934/dcdsb.2017183, 2017.

\bibitem[Xiang et~al.(2003)Xiang, Cheng, Srolovitz, and E]{Xiang2003}
Y.~Xiang, L.~T. Cheng, D.~J. Srolovitz, and W.~E.
\newblock A level set method for dislocation dynamics.
\newblock \emph{Acta Mater.}, 51:\penalty0 5499--5518, 2003.

\bibitem[Yamanaka et~al.(2017)Yamanaka, McReynold, and Voorhees]{Voorhees2017}
A.~Yamanaka, K.~McReynold, and P.~W. Voorhees.
\newblock Phase field crystal simulation of grain boundary motion, grain
  rotation and dislocation reactions in a bcc bicrystal.
\newblock \emph{Acta Mater.}, 133:\penalty0 160--171, 2017.

\bibitem[Zhang et~al.(2005)Zhang, Upmanyu, and Srolovitz]{Zhang2005}
H.~Zhang, M.~Upmanyu, and D.J. Srolovitz.
\newblock Curvature driven grain boundary migration in aluminum: molecular
  dynamics simulations.
\newblock \emph{Acta Mater.}, 53:\penalty0 79--86, 2005.

\bibitem[Zhang et~al.(2017{\natexlab{a}})Zhang, Gu, and Xiang]{Zhang2017}
L.~C. Zhang, Y.~J. Gu, and Y.~Xiang.
\newblock Energy of low angle grain boundaries based on continuum dislocation
  structure.
\newblock \emph{Acta Mater.}, 126:\penalty0 11--24, 2017{\natexlab{a}}.

\bibitem[Zhang et~al.(2017{\natexlab{b}})Zhang, Han, Xiang, and
  Srolovitz]{ZhangLuchan2}
L.~C. Zhang, J.~Han, Y.~Xiang, and D.~J. Srolovitz.
\newblock The equation of motion for a grain boundary.
\newblock \emph{Phys. Rev. Lett.}, 119:\penalty0 246101, 2017{\natexlab{b}}.

\bibitem[Zhu and Xiang(2010)]{Xiang2010}
X.~H. Zhu and Y.~Xiang.
\newblock Continuum model for dislocation dynamics in a slip plane.
\newblock \emph{Philos. Mag.}, 90:\penalty0 4409--4428, 2010.

\bibitem[Zhu and Xiang(2014)]{Zhu2014175}
X.~H. Zhu and Y.~Xiang.
\newblock Continuum framework for dislocation structure, energy and dynamics of
  dislocation arrays and low angle grain boundaries.
\newblock \emph{J. Mech. Phys. Solids}, 69:\penalty0 175 -- 194, 2014.

\end{thebibliography}
\end{document}